\documentclass[]{tech_report_main/fairmeta}
\usepackage{graphicx}%
\graphicspath{{figures/}}%

\usepackage{booktabs}
\usepackage{wrapfig}
\usepackage{tabularx}
\usepackage{textcomp}
\usepackage{stfloats}
\usepackage{url}
\usepackage{verbatim}
\usepackage{titlesec}
\usepackage{tocloft}
\usepackage{adjustbox}
\usepackage{tikz}
\usepackage{comment}
\usepackage{amsmath,amssymb}
\usepackage{colortbl}
\usepackage{color}
\usepackage{hyperref}
\usepackage{graphicx}
\usepackage{subcaption} 
\usepackage{multirow}
\usepackage{booktabs}
\usepackage{xcolor}
\usepackage{pifont}
\usepackage{bbding}
\usepackage{fontawesome}
\usepackage{xspace}
\usepackage{float}
\usepackage{algorithm}
\usepackage{listings}
\usepackage{algcompatible}
\usepackage{algpseudocode}
\usepackage{soul}
\usepackage{array}
\usepackage{tcolorbox}
\usepackage{diagbox}
\usepackage{makecell}
\usepackage{rotating}
\usepackage{enumitem}

\usepackage{amsmath,amsfonts,bm}

\def\eqref#1{equation~\ref{#1}}

\def\1{\bm{1}}

\def\vx{{\bm{x}}}
\def\vy{{\bm{y}}}
\def\vz{{\bm{z}}}

\DeclareMathAlphabet{\mathsfit}{\encodingdefault}{\sfdefault}{m}{sl}
\SetMathAlphabet{\mathsfit}{bold}{\encodingdefault}{\sfdefault}{bx}{n}

\RequirePackage{xspace}
\makeatletter
\DeclareRobustCommand\onedot{\futurelet\@let@token\@onedot}
\def\@onedot{\ifx\@let@token.\else.\null\fi\xspace}

\makeatother

\definecolor{adptorange}{RGB}{248, 205, 172}
\definecolor{cmpblue}{RGB}{189, 215, 238}
\definecolor{cmpblue}{RGB}{189, 215, 238}

\definecolor{our_red}{RGB}{232,157,160}
\definecolor{our_blue}{RGB}{136,206,230}
\definecolor{our_orange}{RGB}{246,200,168}
\definecolor{our_green}{RGB}{178,211,164}

\definecolor{attn_code0}{RGB}{247,215,200}
\definecolor{attn_code1}{RGB}{238,169,139}
\definecolor{mlp_code0}{RGB}{204,201,221}
\definecolor{mlp_code1}{RGB}{102,95,153}

\definecolor{token_blue}{RGB}{84, 120, 140}

\def\vx{{\bm{x}}}

\newlength\savewidth

\newcolumntype{x}[1]{>{\centering\arraybackslash}p{#1pt}}
\newcolumntype{y}[1]{>{\raggedright\arraybackslash}p{#1pt}}
\newcolumntype{z}[1]{>{\raggedleft\arraybackslash}p{#1pt}}

\renewcommand{\paragraph}[1]{\vspace{1mm}\noindent\textbf{#1}}

\renewcommand{\paragraph}[1]{\vspace{1.25mm}\noindent\textbf{#1}}

\definecolor{codeblue}{rgb}{0.25, 0.5, 0.5}
\definecolor{codekw}{rgb}{0.35, 0.35, 0.75}
\lstdefinestyle{Pytorch}{
    language = Python,
    backgroundcolor = \color{white},
    basicstyle = \fontsize{9pt}{8pt}\selectfont\ttfamily\bfseries,
    columns = fullflexible,
    aboveskip=1pt,
    belowskip=1pt,
    breaklines = true,
    captionpos = b,
    commentstyle = \color{codeblue},
    keywordstyle = \color{codekw},
}

\definecolor{green}{HTML}{009000}
\definecolor{red}{HTML}{ea4335}

\newcommand{\eqcontrib}{\clubsuit}
\newcommand{\correspond}{\spadesuit}

\newcommand{\methodabbr}{Lingshu-Cell\xspace}

\newcommand{\myparagraph}[1]{\vspace{0.5\baselineskip}\noindent\textbf{#1}}

\makeatletter
\def\botrule{\noalign{\ifnum0=`}\fi
  \hrule \@height 3pt \@width 0pt
  \hrule \@height 0.75\p@ %
  \hrule \@height 3pt \@width 0pt
  \futurelet\@tempa\@xhline}
\makeatother

\newtcolorbox{promptblock}{
    colback=gray!5,
    colframe=gray!15,
    boxrule=0.5pt,
    arc=3pt,
    left=12pt,
    right=12pt,
    top=8pt,
    bottom=8pt,
    boxsep=8pt,
    breakable
}

\title{\methodabbr: A generative cellular world model for transcriptome modeling toward virtual cells}

\author[\eqcontrib]{Han Zhang}
\author[\eqcontrib]{Guo-Hua Yuan}
\author[\eqcontrib]{Chaohao Yuan}
\author{Tingyang Xu}
\author{Tian Bian}
\author{Hong Cheng}
\author{Wenbing Huang}
\author[\correspond]{Deli Zhao}
\author[\correspond]{Yu Rong}

\affiliation{DAMO Academy, Alibaba Group}

\contribution[\eqcontrib]{Equal Contribution}
\contribution[\correspond]{Project Leader}

\abstract{

Modeling cellular states and predicting their responses to perturbations are central challenges in computational biology and the development of virtual cells. 
Existing foundation models for single-cell transcriptomics provide powerful static representations, but they do not explicitly model the distribution of cellular states for generative simulation.
Here, we introduce Lingshu-Cell, a masked discrete diffusion model that learns transcriptomic state distributions and supports conditional simulation under perturbation. 
By operating directly in a discrete token space that is compatible with the sparse, non-sequential nature of single-cell transcriptomic data, Lingshu-Cell captures complex transcriptome-wide expression dependencies  across approximately 18,000 genes without relying on prior gene selection, such as filtering by high variability or ranking by expression level.
Across diverse tissues and species, Lingshu-Cell accurately reproduces transcriptomic distributions, marker-gene expression patterns and cell-subtype proportions, demonstrating its ability to capture complex cellular heterogeneity.
Moreover, by jointly embedding cell type or donor identity with perturbation, Lingshu-Cell can predict whole-transcriptome expression changes for novel combinations of identity and perturbation. 
It achieves leading performance on the Virtual Cell Challenge H1 genetic perturbation benchmark and in predicting cytokine-induced responses in human PBMCs. Together, these results establish Lingshu-Cell as a flexible cellular world model for \emph{in silico} simulation of cell states and perturbation responses, laying the foundation for a new paradigm in biological discovery and perturbation screening.
}

\date{\today}
\correspondence{Deli Zhao and Yu Rong, \texttt{\{deli.zdl, royrong.ry\}@alibaba-inc.com}}

\begin{document}
\thispagestyle{firstheader}
\maketitle
\pagestyle{fancy}
\fancyhf{}
\fancyfoot[C]{\thepage}

\section{Introduction}\label{sec:intro}

Over the past decade, the rapid expansion of large-scale single-cell RNA sequencing (scRNA-seq) datasets has enabled increasingly comprehensive characterization of cell states across diverse tissues, species, and physiological conditions. Yet most analyses built on these atlases remain primarily descriptive, focusing on annotation, clustering, and comparative characterization rather than predictive modeling. A central challenge is therefore to develop computational frameworks that can capture the distribution of cellular states, generate realistic cellular heterogeneity, and simulate how cells respond to perturbation. Developing such generative capacities would unlock profound biological applications, empowering researchers to conduct large-scale in silico experiments to dissect disease mechanisms, screen potential therapeutics, and map complex developmental trajectories.

To encapsulate this overarching goal, we formally conceptualize such a comprehensive framework as a \emph{cellular world model}. Analogous to world models in artificial intelligence that learn compact representations of an environment and support conditional simulation, a cellular world model aims to represent the distribution of transcriptomic states and their conditional dynamics. By explicitly modeling this intrinsic state space, such systems can move single-cell biology beyond static cataloging toward \emph{in silico} environments capable of simulating high-fidelity cellular states and their responses under intervention.

Inspired by the success of foundation models in natural language processing, recent advances in large-scale self-supervised learning for transcriptomics, including scGPT~\citep{cui2024scgpt}, Geneformer~\citep{theodoris2023transfer}, scFoundation~\citep{hao2024large}, and CellFM~\citep{zeng2025cellfm} have shown that pretrained foundation models can capture transferable structure in gene expression and organize cells in shared representation spaces across diverse datasets.  However, these models are primarily optimized for learning static representations rather than generative simulation. Existing generative approaches, such as  scDiffusion~\citep{luo2024scdiffusion} and scVI~\citep{lopez2018deep}, have shown promise for transcriptome generation and perturbation modeling. However, their performance is limited by continuous data assumptions that misalign with the sparse, discrete, and non-sequential nature of single-cell transcriptome data. In parallel, perturbation-focused methods, such as STATE~\citep{adduri2025predicting}, CellFlow~\citep{klein2025cellflow},  scDFM~\citep{yu2026scdfm},  and AlphaCell~\citep{chuai2026towards}, commonly learn direct mappings from control states and perturbation conditions to perturbed outcomes. While effective for specific prediction tasks, these methods do not model the underlying distribution of transcriptomic states or their conditional dynamics. Together, these limitations highlight the need for a cellular world model that explicitly represents transcriptomic state space and supports conditional simulation under perturbation.

Here, we present Lingshu-Cell, a masked discrete diffusion model for transcriptome-wide generative modeling of cellular states.
Lingshu-Cell is trained with a masking-and-prediction objective over discrete gene-expression tokens. This design enables non-autoregressive, bidirectional refinement of whole-transcriptome profiles, while remaining compatible with the sparse, non-sequential nature of scRNA-seq data. Lingshu-Cell directly models transcriptome-wide expression across approximately 18,000 genes without requiring prior gene selection,  such as filtering by high variability or ranking by expression level,
and captures complex combinatorial gene-expression patterns underlying cellular heterogeneity. Across large-scale single-cell datasets spanning nine tissues and five species, Lingshu-Cell reproduces the 
transcriptomic distributions, marker-gene expression patterns and cell-subtype proportions of real scRNA-seq data, enabling realistic simulation of heterogeneous cell populations. Furthermore, Lingshu-Cell embeds cell type  or donor identity with  perturbation context (such as genetic or cytokine perturbation) into a joint latent space for modeling whole-transcriptome expression changes to perturbations.  It achieves leading performance on the Virtual Cell Challenge H1 genetic perturbation benchmark~\citep{roohani2025virtual} using only approximately 0.6 million training cells, and demonstrates strong results on cytokine perturbation prediction in human PBMCs. Together, these results position Lingshu-Cell as a flexible cellular world model for virtual cell modeling and \emph{in silico} perturbation analysis across diverse biological contexts, laying the foundation for a new paradigm in biological discovery and perturbation screening.

\section{Results}\label{sec:results}

\subsection{Overview of the Lingshu-Cell framework}\label{subsec:overview}

\begin{figure}[t]
\centering
\includegraphics[width=0.9\linewidth]{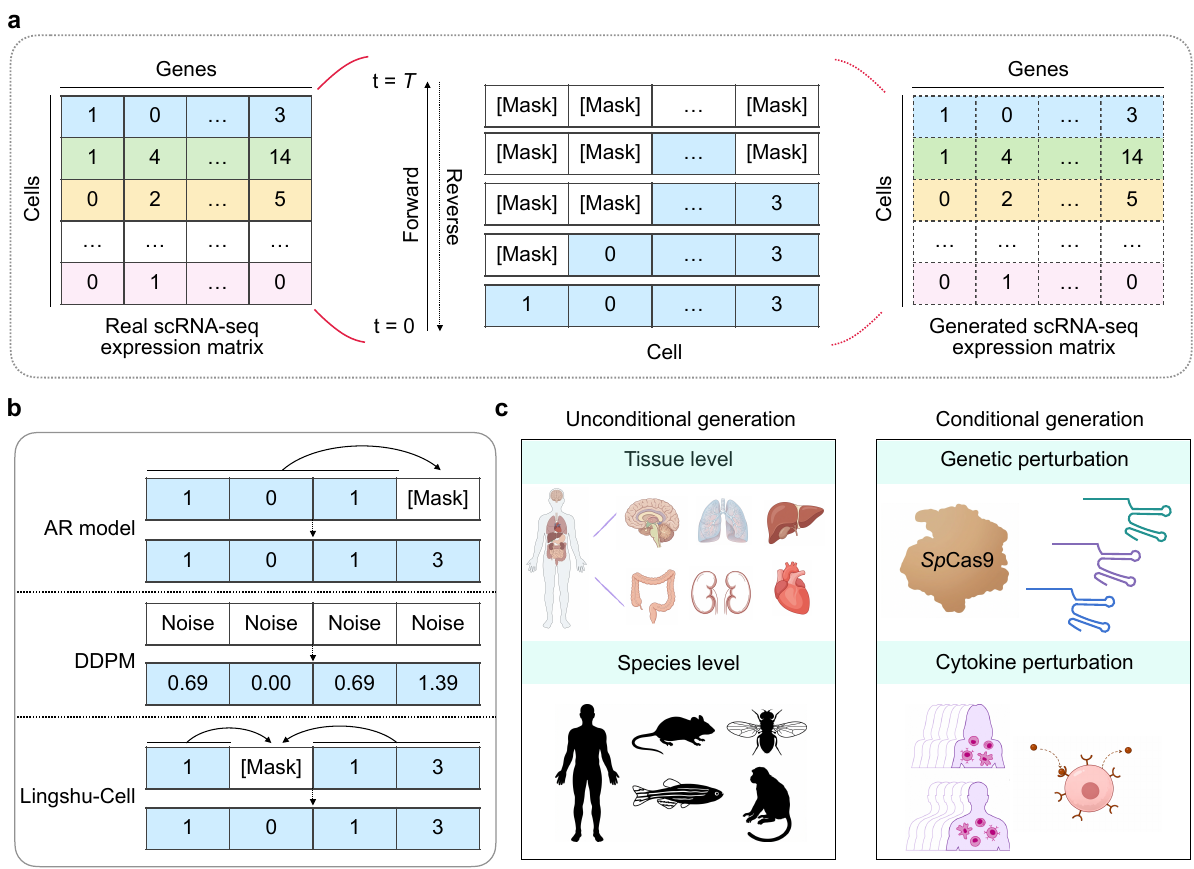}
\caption{\textbf{Overview of the Lingshu-Cell framework.}
\textbf{a}, Lingshu-Cell employs a masked discrete diffusion model to learn and generate single-cell transcriptomic data. In the forward process, gene expression values are progressively masked (from $t = 0$ to $t = T$); in the reverse process, the model iteratively predicts masked values to generate realistic scRNA-seq expression profiles.
\textbf{b}, Comparison of generative paradigms. Unlike autoregressive (AR) models that rely on a fixed sequential order and denoising diffusion probabilistic models (DDPMs) that corrupt all positions with continuous noise, Lingshu-Cell randomly masks and predicts gene expression values in an order-independent manner, which is inherently compatible with the orderless structure of gene expression data.
\textbf{c}, Application scenarios of Lingshu-Cell, including unconditional generation across diverse human tissues and species, and conditional generation for genetic perturbation and cytokine perturbation response prediction.}
\label{fig:figure1}
\end{figure}

To comprehensively model gene expression and characterize cellular states at single-cell resolution, we developed Lingshu-Cell, a novel generative framework for single-cell transcriptomic data based on a masked discrete diffusion model architecture. Specifically, given a real scRNA-seq expression matrix, Lingshu-Cell operates through two coupled processes: in the forward process, gene expression values of each cell are progressively masked from the original observed state ($t=0$) to a fully masked state ($t=T$); in the reverse process, the model iteratively predicts the masked gene expression values, ultimately generating biologically realistic scRNA-seq profiles (Fig.~\ref{fig:figure1}a and Appendix Fig.~\ref{fig:architecture}). This masking-and-prediction paradigm enables Lingshu-Cell to learn complex gene regulatory dependencies while naturally accommodating the orderless structure of gene expression profiles. Accordingly, it eliminates the need for an arbitrary generation order required by AR models and avoids the global continuous-noise corruption used in DDPMs~\citep{ho2020denoising} (Fig.~\ref{fig:figure1}b), which is poorly matched to the discrete and often highly sparse nature of raw scRNA-seq counts. Leveraging this design, we apply Lingshu-Cell to unconditional generation to simulate transcriptomic profiles across diverse human tissues and species, and to conditional generation to predict cellular responses to genetic and cytokine perturbations (Fig.~\ref{fig:figure1}c), moving toward a practical virtual cell model.

\subsection{Lingshu-Cell enables accurate simulation of cell states across diverse species and tissues}\label{subsec:unconditional-generation}

\begin{figure}[t]
\centering
\includegraphics[width=\linewidth]{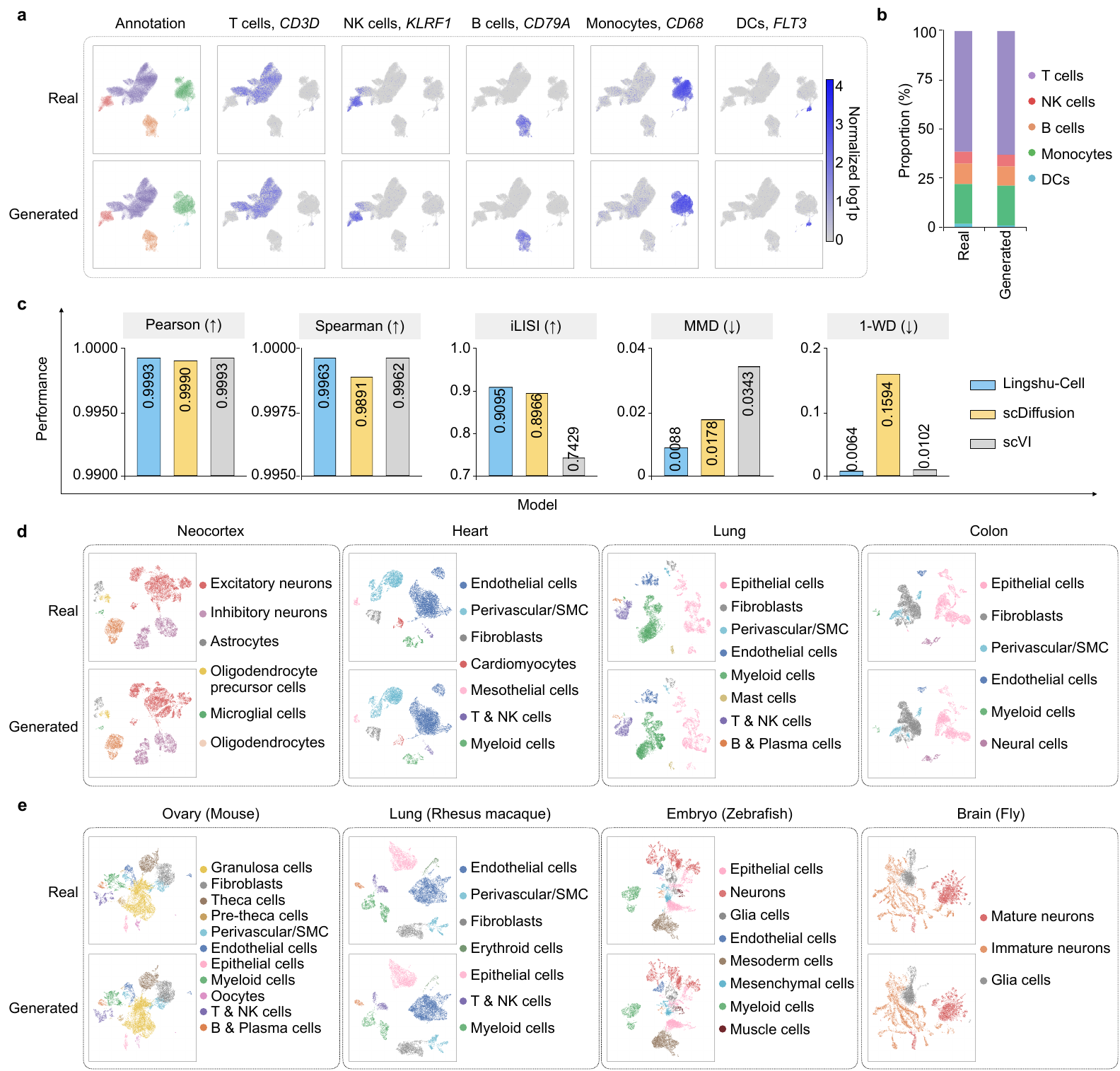}
\caption{\textbf{Unconditional generation of cell states across diverse species and tissues by Lingshu-Cell.}
\textbf{a}, UMAP visualization of real and generated cells (10,000 each, randomly sampled) from the PARSE-PBMC dataset, colored by cell type annotation (left) and normalized expression (log1p) of canonical marker genes for each cell type.
\textbf{b}, Comparison of cell type proportions between real and generated data.
\textbf{c}, Quantitative benchmark comparing Lingshu-Cell, scDiffusion and scVI across five metrics (Pearson correlation, Spearman correlation, MMD, 1-WD and iLISI) on the PARSE-PBMC dataset.
\textbf{d}, Unconditional generation results across human tissues, including neocortex, heart, lung and colon, with UMAP plots showing real (top) and generated (bottom) cells colored by cell type.
\textbf{e}, Unconditional generation results across multiple species, including mouse, rhesus macaque, zebrafish and fly.}
\label{fig:figure2}
\end{figure}

To validate the fundamental capacity of Lingshu-Cell to model cellular gene expression, we first trained the model on the PBS control subset of the PARSE 10M PBMC dataset (629,701 cells) and then randomly generated 10,000 cells, a scale comparable to that of a typical scRNA-seq experiment. By comparing real and generated data, we found that Lingshu-Cell faithfully recapitulated marker-gene expression patterns across the five major PBMC lineages, T cells, NK cells, B cells, monocytes and dendritic cells (Fig.~\ref{fig:figure2}a). The generated cell-type proportions were also highly consistent with those in the real dataset (Fig.~\ref{fig:figure2}b). To reduce potential sampling variability due to the relatively small number of generated cells, we further scaled generation to 200,000 cells. As expected, both marker-gene expression patterns (Appendix Fig.~\ref{fig:ext-data-fig1}a) and cell-type proportions (Appendix Fig.~\ref{fig:ext-data-fig1}b) remained highly concordant with the real data. At this larger scale, we performed higher-resolution annotation and further subdivided PBMCs into 17 subtypes (Appendix Fig.~\ref{fig:ext-data-fig1}c). The generated and real data continued to align closely, indicating that Lingshu-Cell can robustly simulate cellular gene expression at both standard and very large scales.

\FloatBarrier
\FloatBarrier
\begin{figure}[H]
\captionsetup{type=table}
\centering
\caption{Unconditional generation performance of \methodabbr across human tissues and non-human species.}\label{tab:uncond-tissue-species}
\footnotesize
\begin{tabular}{@{}llccccc@{}}
\toprule
& Tissue & Pearson ($\uparrow$) & Spearman ($\uparrow$) & MMD ($\downarrow$) & iLISI ($\uparrow$) & 1-WD ($\downarrow$) \\
\midrule
\multicolumn{7}{@{}l}{{Human tissues (CZ CELLxGENE)}} \\
& Neocortex & 0.9995 & 0.9991 & 0.0128 & 0.9053 & 0.0105 \\
& Heart & 0.9992 & 0.9987 & 0.0196 & 0.8972 & 0.0096 \\
& Lung & 0.9967 & 0.9970 & 0.0314 & 0.8906 & 0.0159 \\
& Colon & 0.9966 & 0.9960 & 0.0376 & 0.8815 & 0.0152 \\
\midrule
\multicolumn{7}{@{}l}{{Non-human species - Tissues}} \\
& Mouse - Ovary & 0.9996 & 0.9989 & 0.0116 & 0.9011 & 0.0077 \\
& Rhesus macaque - Lung & 0.9985 & 0.9970 & 0.0218 & 0.8926 & 0.0149 \\
& Zebrafish - Embryo & 0.9983 & 0.9974 & 0.0143 & 0.9035 & 0.0089 \\
& Fly - Brain & 0.9984 & 0.9929 & 0.0163 & 0.8876 & 0.0107 \\
\botrule
\end{tabular}

\end{figure}
\vspace{-10pt}
We quantified these observations by benchmarking Lingshu-Cell against scDiffusion~\citep{luo2024scdiffusion} and scVI~\citep{lopez2018deep} on the PBMC dataset using five complementary metrics, including Pearson and Spearman correlations to assess expression concordance, as well as maximum mean discrepancy (MMD), gene-averaged 1-Wasserstein distance (1-WD), and integration local inverse Simpson's index (iLISI) to evaluate distributional similarity and integration quality (Fig.~\ref{fig:figure2}c). All three methods achieved uniformly high gene expression correlations, suggesting that they were all able to capture global gene expression patterns. By contrast, MMD, 1-WD, and iLISI revealed clearer differences in generative quality. In particular, Lingshu-Cell achieved the lowest MMD (0.0088, compared with 0.0178 for scDiffusion and 0.0343 for scVI), indicating the closest overall match between generated and real expression distributions, consistent with the trends observed in cell-level UMAP visualizations and cell-type proportion analyses. These results further suggest that, by achieving the best performance across all five metrics, Lingshu-Cell provides the most faithful modeling of the PBMC scRNA-seq dataset.

To further assess generalizability across tissues and minimize dataset-specific effects, we assembled 2,602,318 cells from the CZ CELLxGENE database~\citep{czi2025cz} spanning eight human tissues (neocortex, thymus, heart, lung, liver, colon, kidney and breast). Quality control and summary statistics revealed substantial heterogeneity across tissues and batches, including large differences in cell numbers, detected genes, total counts and the percentage of mitochondrial reads (Appendix Fig.~\ref{fig:ext-data-fig2}). Despite this variability, Lingshu-Cell consistently produced high-quality samples that accurately captured major cell types as well as tissue-specific cell types in each tissue (Fig.~\ref{fig:figure2}d, Appendix Fig.~\ref{fig:ext-data-fig3} and Table~\ref{tab:uncond-tissue-species}).

Moreover, we extended Lingshu-Cell to single-cell datasets from four additional species, totaling 247,899 cells across diverse tissues, including mouse ovary, rhesus macaque lung, zebrafish embryo, and fly brain. Although these datasets also exhibited pronounced differences in quality-control metrics and data distributions (Appendix Fig.~\ref{fig:ext-data-fig4}), Lingshu-Cell accurately generated the corresponding cell types with high fidelity (Fig.~\ref{fig:figure2}e and Table~\ref{tab:uncond-tissue-species}). 

Together, these results established that Lingshu-Cell generalizes reliably in the unconditional setting across tissues and species, providing a foundation for evaluating its performance under controlled perturbations.

\subsection{Lingshu-Cell accurately predicts single-cell transcriptomic responses to genetic perturbations in cell lines}

\begin{figure}[t]
\centering
\includegraphics[width=0.95\linewidth]{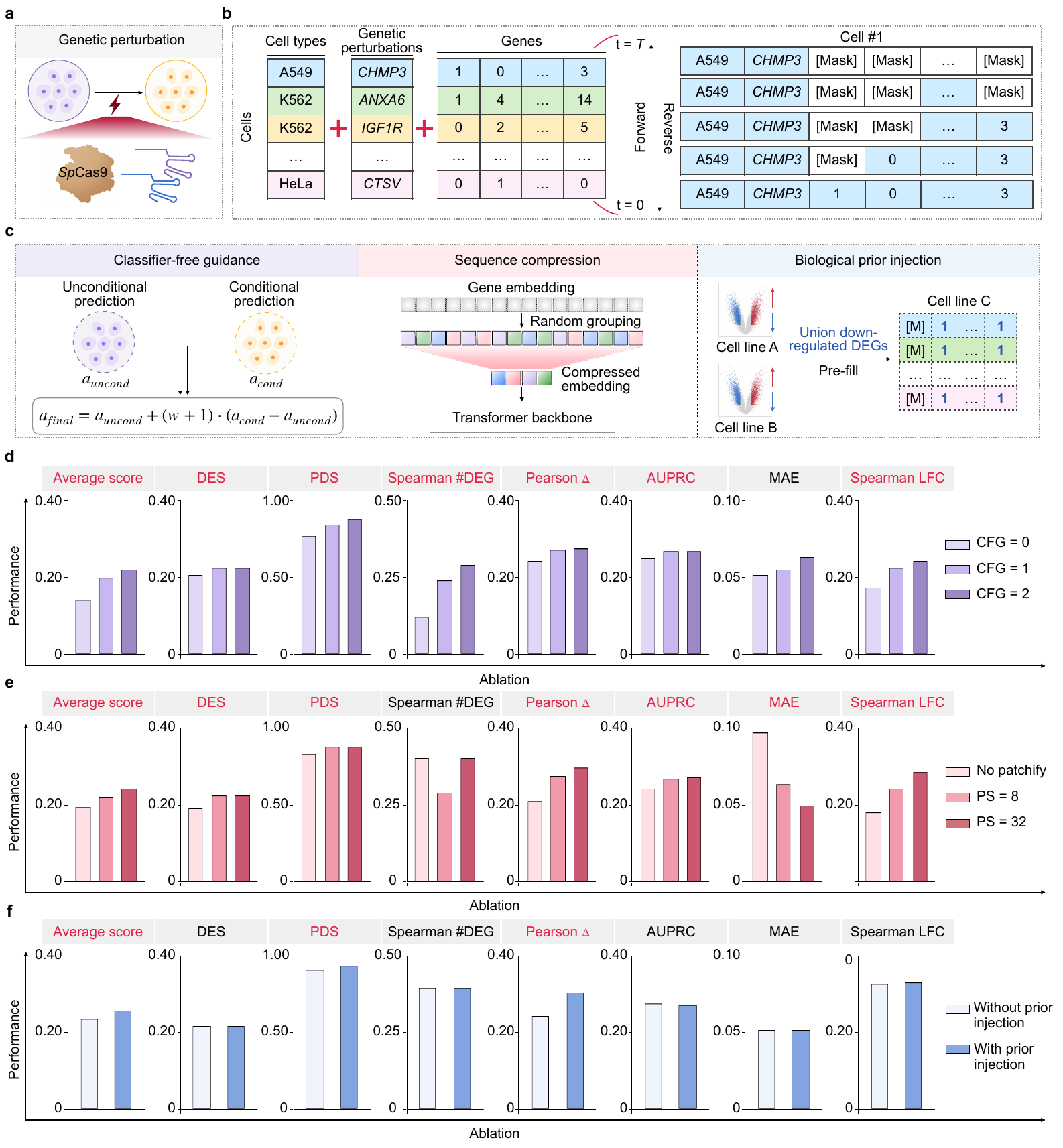}
\caption{\textbf{Accurate prediction of single-cell transcriptomic responses to genetic perturbations in cell lines by Lingshu-Cell.}
\textbf{a}, Schematic of CRISPR-based genetic perturbation and the resulting transcriptomic changes.
\textbf{b}, Conditional generation framework for perturbation prediction. Cell type and perturbation target are provided as conditioning inputs, and a masked diffusion model iteratively predicts gene expression values to generate perturbation-specific expression profiles.
\textbf{c}, Three design components of Lingshu-Cell: classifier-free guidance (CFG), sequence compression, and biological prior injection (see Methods).
\textbf{d}, Ablation study of CFG guidance weight. Bar plots show prediction performance across eight metrics (DES, PDS, MAE, Spearman \#DEG, Spearman LFC, AUPRC, Pearson-$\Delta$, and average score) on the H1 test set ($n = 100$ perturbation targets).
\textbf{e}, Ablation study of sequence compression, comparing uncompressed input with patch sizes of 8 and 32.
\textbf{f}, Ablation study of biological prior injection, comparing prediction performance with and without prior injection.
In \textbf{d--f}, metrics highlighted in red denote improved performance relative to the corresponding baseline in each ablation setting.}
\label{fig:figure3}
\end{figure}

Given the strong performance of \methodabbr in modeling cellular gene expression distributions in the unconditional setting, we next asked whether the same framework could support conditional generation of genetic perturbation responses (Fig.~\ref{fig:figure3}a). Because \methodabbr operates directly in a discrete token space, cell-type identity and perturbation-target information can be introduced as additional tokens prepended to the expression sequence, enabling conditional generation within a unified modeling framework (Fig.~\ref{fig:figure3}b). This formulation allows the model to exploit shared perturbation-response patterns across cell types and generalize to previously unseen combinations of cell type and perturbation target.

We evaluated conditional generation on the H1 genetic perturbation dataset from the Virtual Cell Challenge (VCC)~\citep{roohani2025virtual}. The training data comprised perturbation profiles from external cell lines for all perturbations whose target genes overlapped the 300 perturbation targets defined in the H1 dataset (n = 323,913 cells), together with H1 cells from the 150 training targets (n = 183,097 cells). Unperturbed control
\FloatBarrier
\begin{figure}[H]
\captionsetup{type=table}
\centering
\caption{Genetic perturbation prediction on the VCC leaderboard~\citep{roohani2025virtual}. Teams are ordered by final ranking. Avg Rank: average rank across the top 25 teams. Best per column in bold. See \Cref{tab:vcc-leaderboard-full} for the full top-25 ranking.}\label{tab:vcc-leaderboard}
\footnotesize
\setlength{\tabcolsep}{3pt}
\begin{tabular*}{\textwidth}{@{\extracolsep\fill}lcccccccc}
\toprule
Team & Avg Rank$\downarrow$ & DES$\uparrow$ & PDS$\uparrow$ & MAE$\downarrow$ & Sp. \#DEG$\uparrow$ & Sp.\ LFC$\uparrow$ & AUPRC$\uparrow$ & Pearson-$\Delta$$\uparrow$ \\
\midrule
\methodabbr & \textbf{8.7} & 0.216 & 0.748 & \textbf{0.052} & 0.394 & 0.331 & 0.272 & \textbf{0.306} \\
cleopatra & 9.1 & 0.228 & 0.747 & 0.086 & 0.473 & \textbf{0.396} & 0.266 & 0.203 \\
xBio & 10.7 & 0.305 & \textbf{0.811} & 0.770 & \textbf{0.564} & 0.087 & 0.252 & 0.217 \\
Cellock Holmes & 10.9 & \textbf{0.356} & 0.679 & 0.239 & 0 & 0.238 & 0.576 & 0.125 \\
Shippers & 11.0 & 0.354 & 0.699 & 0.231 & 0 & 0.227 & 0.576 & 0.123 \\
Mean Predictors & 11.4 & 0.305 & 0.741 & 6.723 & 0.294 & 0.213 & \textbf{0.582} & 0.217 \\
\botrule
\end{tabular*}

\end{figure}
\vspace{-10pt}
\FloatBarrier
\noindent cells were also included in the training set to enable classifier-free guidance~\citep{ho2021classifierfree} (see Methods \Cref{subsec:cfg} and Appendix \Cref{subsec:training-details} for details). Model performance was evaluated on H1 cells from 50 validation targets (n = 60,751 cells) and 100 test targets (n = 132,670 cells) that were held out during training. To improve prediction accuracy, we incorporated three strategies targeting complementary aspects of conditional generation. First, we applied classifier-free guidance (CFG) to steer sampling toward transcriptomic states more consistent with the perturbed condition, thereby improving the fidelity of generated perturbation responses (Fig.~\ref{fig:figure3}c, left). Second, we adopted sequence compression to transform the high-dimensional gene expression sequence into a shorter sequence of embeddings with higher information density, improving modeling efficiency while facilitating the capture of global expression patterns (Fig.~\ref{fig:figure3}c, middle). Third, we introduced biological prior projection, in which perturbation-responsive genes were identified from external cell lines by first determining affected genes within each cell line and then taking their union to form a perturbation-specific prior gene set. This prior set was preferentially used to initialize masked positions at the start of generation, thereby injecting biologically informed prior knowledge into the sampling process (Fig.~\ref{fig:figure3}c, right).

We then performed ablation experiments to quantify the contribution of each component individually. As expected, all three strategies yielded measurable performance gains (Fig.~\ref{fig:figure3}d). In particular, removing CFG led to poorer results, especially on perturbation direction similarity and correlation based metrics, consistent with its role in biasing generation toward the perturbed expression manifold. Among the tested guidance strengths, $\mathrm{CFG}=2$ achieved the best overall performance (Fig.~\ref{fig:figure3}e). Sequence compression also had a substantial effect: a patch size of 32 outperformed smaller patch sizes in both average score and Spearman \#DEG correlation, reaching 0.405 compared with 0.292 for a patch size of 8 (Fig.~\ref{fig:figure3}f), indicating that moderate compression improves representation of high dimensional gene expression signals. Incorporating biological priors further improved perturbation direction similarity and Pearson-$\Delta$ correlation (Fig.~\ref{fig:figure3}g), supporting the value of incorporating perturbation priors aggregated across external cell lines into the generation process.

After integrating all three strategies, \methodabbr achieved its best overall performance on the VCC H1 test set. We further compared the full model with the top performing published methods from the Virtual Cell Challenge (see Methods). Across seven evaluation metrics, \methodabbr obtained the best average rank (Table~\ref{tab:vcc-leaderboard}), indicating the most consistent overall performance among all compared methods. It achieved the lowest MAE (0.052) and the highest Pearson-$\Delta$ correlation (0.306). Although other methods ranked first on individual metrics, \methodabbr provided the best overall balance across all seven evaluation criteria.

Together, these results demonstrate that \methodabbr, as a general-purpose generative model, can effectively predict transcriptomic responses to genetic perturbations and outperform task-specific predictive models on a standard benchmark.

\subsection{Lingshu-Cell accurately predicts single-cell transcriptomic responses to cytokine perturbations in PBMCs}

\begin{figure}[t]
\centering
\includegraphics[width=\linewidth]{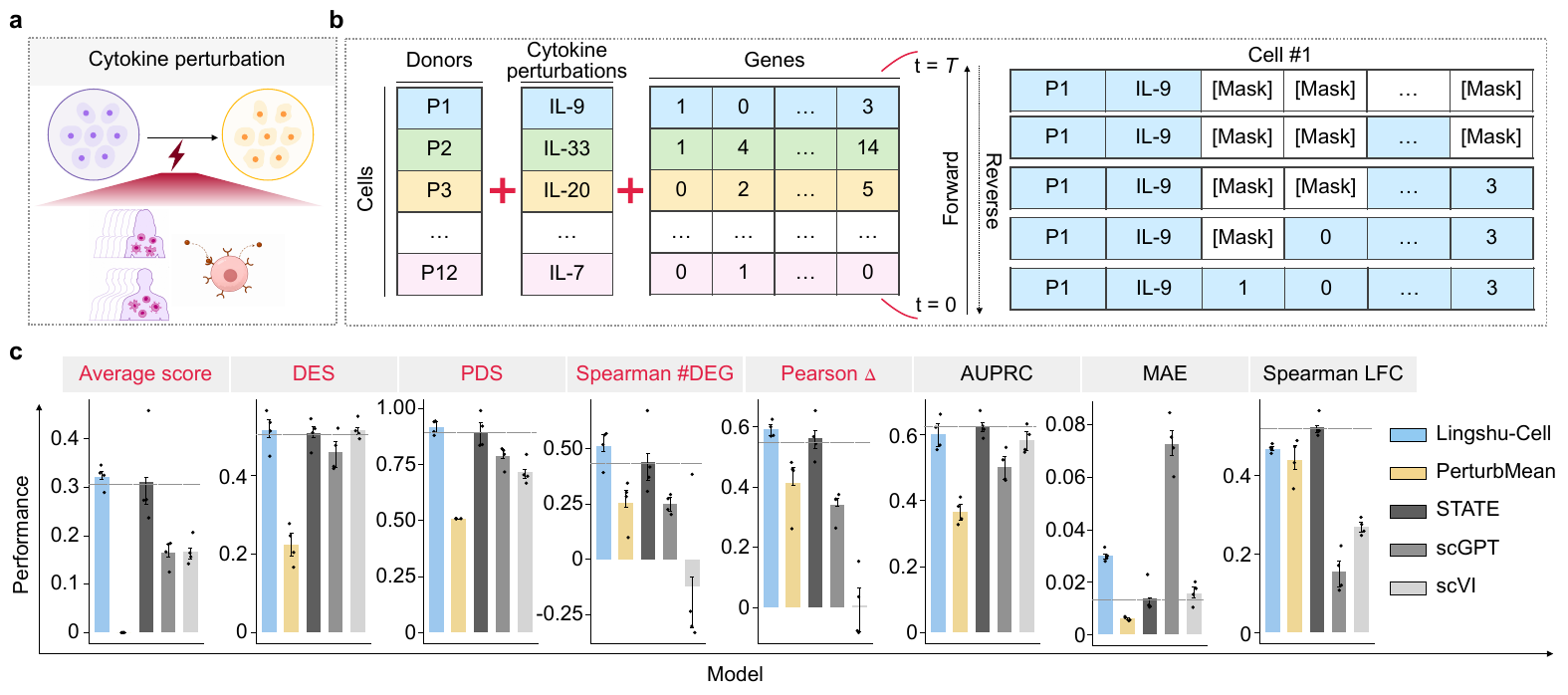}
\caption{\textbf{Accurate prediction of single-cell transcriptomic responses to cytokine perturbations in PBMCs by Lingshu-Cell.}
\textbf{a}, Schematic of cytokine-induced transcriptomic perturbation.
\textbf{b}, Conditional generation framework for cytokine perturbation prediction. Donor identity and cytokine condition are provided as conditioning inputs, and a masked diffusion process iteratively predicts gene expression values to generate perturbation-specific expression profiles.
\textbf{c}, Prediction performance of Lingshu-Cell, PertMean, STATE, scGPT, and scVI on the PARSE 10M PBMC dataset, evaluated across eight metrics (as defined in Fig. 3d). Bar plots show performance on a test set comprising 4 of 12 donors, with 70\% of cytokine conditions (63 of 90) held out for each donor. Bars indicate mean performance across donors; error bars, interquartile range (Q1--Q3); points, individual donor values. Metrics for which Lingshu-Cell achieves the best performance among all methods are highlighted in red.}
\label{fig:figure4}
\end{figure}

Having demonstrated strong performance on genetic perturbation prediction in the cell-line system, we next investigated whether \methodabbr could extrapolate to a distinct perturbation modality and a higher level of biological complexity. We therefore evaluated \methodabbr on cytokine-driven transcriptomic perturbations in the PARSE 10M PBMC dataset, which profiles peripheral blood mononuclear cells (PBMCs) from 12 donors, each exposed to 90 distinct cytokine conditions alongside an unperturbed (PBS) control (Fig.~\ref{fig:figure4}a). As in the genetic perturbation setting, conditional generation was implemented by prepending condition tokens to the expression sequence; here, donor identity and cytokine condition were introduced as additional tokens, enabling the model to generate transcriptomic responses conditioned jointly on donor context and stimulation type (Fig.~\ref{fig:figure4}b). Unlike the previous benchmark based on genetic perturbations in cell lines, this task requires modeling signaling-induced responses in donor-derived immune cells across individuals. To assess generalization, we randomly selected 4 of the 12 donors and, for each donor, held out 70\% of cytokine conditions (63 of 90) as the test set.

\methodabbr achieved the highest average score across all evaluated methods (Fig.~\ref{fig:figure4}c). At the level of overall expression profiles, it ranked first in both PDS and Pearson-$\Delta$ correlation, indicating that the predicted transcriptomes preserved the distinct identity of each cytokine condition and accurately captured the direction and magnitude of cytokine-induced expression changes. At the differential expression level, \methodabbr also achieved the highest Spearman \#DEG correlation across perturbations, indicating that it correctly recovered the relative strength of transcriptional responses induced by different cytokines. Together with strong performance on DES, Spearman LFC, and AUPRC, these results show that \methodabbr not only identified the genes responsive to each cytokine but also captured the overall scale and structure of their transcriptional effects.

These findings demonstrate that \methodabbr generalizes from genetic perturbations in cell lines to cytokine stimulations in donor-derived PBMCs, supporting its applicability across perturbation modalities and biological contexts. Although genetic perturbations and cytokine stimulations act through fundamentally different mechanisms, \methodabbr achieved leading performance in both settings (Fig.~\ref{fig:figure3}; Fig.~\ref{fig:figure4}).
This consistency highlights the potential of the conditional generation framework as a unified approach for predicting cellular responses to diverse perturbations across experimental contexts.

\section{Discussion}\label{sec12}

\methodabbr{} demonstrates that MDDM can serve as a unified generative framework for single-cell transcriptomics, establishing a computational foundation for a cellular world model. By directly modeling transcriptome-wide expression across approximately 18,000 genes without prior gene selection based on high variability or expression level, \methodabbr{} shifts single-cell foundation models from static representation learning to generative simulation. Within a single architecture, it achieves high-fidelity cell generation across diverse tissues and species, capturing true cellular heterogeneity, while successfully predicting responses to both genetic (VCC) and cytokine perturbations (PARSE). Together, these results mark a critical step toward interactive virtual cells.

This success stems from aligning the computational paradigm with the physical properties of biological data.
By operating in a discrete expression space, masked discrete diffusion model avoids the artificial gene-ordering bias of autoregressive models~\citep{austin2021structured, nie2025large}, the information bottleneck of variational autoencoders~\citep{alemi2017deep}, and the distributional mismatch between continuous noise processes and sparse, discrete count data~\citep{risso2018general, vignac2023digress}, aligning naturally with the permutation invariance and zero-inflated sparsity intrinsic to transcriptomic measurements. 

Despite these strong empirical results, several limitations, however, merit discussion. First, current evaluations rely on population-level distributional metrics (e.g., MMD, iLISI) and pseudobulk correlations, which cannot fully assess single-cell biological plausibility or the preservation of extremely rare states. More fundamentally, high-fidelity generation does not imply biological causality: faithful recapitulation of expression distributions does not necessarily  reflect the causal regulatory mechanisms that produce them. Therefore, \methodabbr{} should currently be regarded as a powerful tool for probabilistic hypothesis generation. Its predictions still require rigorous wet-lab experimental validation. 
Finally, the current model operates exclusively on transcriptomic data; a more complete virtual cell would additionally integrate epigenomic, proteomic, metabolomic, and spatial modalities~\citep{hao2021integrated, argelaguet2020mofa+, dries2021giotto}.

Several promising directions extend naturally from the present study. An important next step is to extend the current framework beyond single-gene perturbations to more complex interventions, including drug-induced, multi-target, and combinatorial interventions, where dose dependence and temporal dynamics are likely to play critical roles~\citep{lotfollahi2019scgen, hetzel2022predicting, bergen2020generalizing}. More broadly, we envision jointly modeling chromatin accessibility, protein abundances, and spatially resolved transcriptomics within the same discrete diffusion framework, and incorporating temporal structure to simulate dynamic cellular trajectories such as differentiation and disease progression. Ultimately, one of the most compelling applications may be closed-loop experimentation, in which model predictions guide targeted perturbations and newly generated data iteratively refine the model itself. Such a framework would move beyond static data fitting toward an adaptive platform for biological discovery. In this sense, \methodabbr{} establishes MDDM as a promising paradigm for modeling cellular behavior, paving the way for a truly predictive cellular world model.

\section{Methods}\label{sec:methods}
We first introduce the general masked diffusion formulation underlying \methodabbr, and then describe how single-cell expression profiles are represented as discrete token sequences for model training and generation. We next present the main methodological components of the framework, including embedding-space sequence compression, conditional generation and inference-time biological prior injection.

\subsection{Preliminaries} \label{subsec:preliminaries}
Here we introduce the preliminary concepts of masked discrete diffusion models (MDDMs) that form the foundation of \methodabbr.

Let $x_0 = [x_0^1, x_0^2, \dots, x_0^L]$ be a fully observed discrete sequence of length $L$, where each token $x_0^i$ belongs to a predefined vocabulary $\mathcal{V}$. To model the joint distribution $p_{\text{data}}(x_0)$ without the left-to-right inductive bias of autoregressive models, MDDMs introduce a forward masking process and a learnable reverse generation process.

\myparagraph{Forward Process.} The forward process gradually and independently masks tokens in $x_0$ over a continuous time variable $t \in [0, 1]$. At $t=0$, the sequence is completely clean; at $t=1$, the sequence is fully masked, representing a sequence of special mask tokens $M \notin \mathcal{V}$. For any intermediate timestep $t$, each token $x_t^i$ in the partially masked sequence $x_t$ is independently replaced by $M$ with probability $t$, or remains the original token with probability $1-t$. Formally, the transition probability is defined as:
\begin{equation}
q_{t|0}(x_t^i|x_0^i) = 
\begin{cases} 
1 - t, & x_t^i = x_0^i, \\ 
t, & x_t^i = M. 
\end{cases}
\end{equation}

\myparagraph{Reverse Process and Training Objective.} The reverse process aims to recover the true data distribution by iteratively predicting the masked tokens as $t$ transitions from $1$ back to $0$. The core of this process is a parametric mask predictor neural network, $p_\theta(\cdot|x_t)$, which takes the partially masked sequence $x_t$ as input and predicts the original tokens for all masked positions simultaneously.

The model is optimized using a cross-entropy loss computed exclusively on the masked tokens:
\begin{equation}
\mathcal{L}(\theta) \triangleq -\mathbb{E}_{t, x_0, x_t} \left[ \frac{1}{t} \sum_{i=1}^L \mathbb{I}[x_t^i = M] \log p_\theta(x_0^i | x_t) \right], \label{eq:loss}
\end{equation}
where $t \sim \mathcal{U}(0, 1)$, and $\mathbb{I}[\cdot]$ is the indicator function. This objective provides a principled variational upper bound on the negative log-likelihood of the true data distribution.

\myparagraph{Inference and Sampling.} During inference, the generation begins with a fully masked sequence $x_1$ of length $L$. The reverse process is discretized into $N$ steps. At an intermediate step transitioning from time $t \in (0, 1]$ to $s \in [0, t)$, the mask predictor $p_\theta(\cdot|x_t)$ estimates all masked tokens. Subsequently, a proportion (specifically $\frac{s}{t}$ in expectation) of the predicted tokens are remasked to construct $x_s$, ensuring the transition aligns with the forward process dynamics.

\subsection{Representing Single-Cell Data as Discrete Sequences} \label{subsec:modeling}

MDDMs operate on discrete token sequences (\Cref{subsec:preliminaries}). Here we describe how each cell's gene expression profile, measured as UMI counts, is mapped to such a discrete sequence, defining the modeling space for both training and generation.

\myparagraph{Single-cell data preparation.}
Let $\mathbf{X}\in\mathbb{Z}_{\ge 0}^{N\times G}$ denote the cell-by-gene UMI count matrix, where $N$ is the number of cells and $G$ is the number of genes. The $i$-th cell is represented by a count vector $\vx^{(i)}=(x^{(i)}_1,\dots,x^{(i)}_G)\in\mathbb{Z}_{\ge 0}^{G}$ over a fixed gene list. 
The UMI count matrix is preprocessed using standard single-cell analysis pipelines; details are deferred to the Appendix.

\myparagraph{Quantization.}
Although UMI counts are discrete, they span a broad dynamic range with a heavy tail, making exact-count tokenization inefficient. We therefore quantized each gene’s count into a finite set of expression levels, reducing the size of the discrete state space while preserving resolution at low counts. We define a shared quantization function with $B$ non-overflow bins and a unified overflow token $\text{OVF}$:
\begin{equation}
q:\mathbb{Z}_{\ge 0}\rightarrow \{0,1,\dots,B-1\}\cup\{\text{OVF}\},
\end{equation}
and map each entry to a bin index $z_g^{(i)} = q(x_g^{(i)})$. The modeled range is capped at $C$. Specifically, we define:
\begin{equation}
q(x)=
\begin{cases}
x, & 0\le x < 100,\\
100 + 90\cdot\max(0, k(x)-2) + r(x), & 100\le x \le C, \\
\text{OVF}, & x > C,
\end{cases}
\label{eq:quantization}
\end{equation}
where 
\begin{equation}
k(x)=\left\lfloor\log_{10}(\max(x,1))\right\rfloor,
\end{equation}
and, for $x\ge 100$,
\begin{equation}
r(x)=\left\lfloor \frac{x-10^{k(x)}}{\Delta(x)}\right\rfloor,
\end{equation}
with
\begin{equation}
\Delta(x)=10^{k(x)-1}.
\label{eq:quantization_step}
\end{equation}
Here, $k(x)$ denotes the decade index of $x$, $r(x)$ denotes the within-decade offset, and $\Delta(x)$ is the adaptive step size at that scale. This construction yields 100 bins for $[0,99]$ and 90 bins per decade for $[10^k,10^{k+1})$ with $k\ge 2$. Thus, the quantization preserves approximately the first two significant digits of $x$.

Unless specified otherwise, we use $C=9999$ throughout; in this case, the non-overflow vocabulary size is $B=100+90+90=280$, and $|\mathcal{V}|=281$ after including the overflow token (excluding special tokens such as \texttt{[MASK]} that are part of the MDDM training setup).

\myparagraph{Discrete sequence representation.}
Finally, each cell is represented as a discrete sequence $\vz^{(i)}\in\mathbb{Z}_{\ge 0}^{G}$ obtained by applying the quantization function element-wise,
\begin{equation}
\vz^{(i)}=\left(q\!\left(x^{(i)}_1\right),\,q\!\left(x^{(i)}_2\right),\,\dots,\,q\!\left(x^{(i)}_G\right)\right),
\label{eq:discrete_sequence}
\end{equation}
which serves as the input/output space of our MDDM for modeling single-cell data as discrete sequences.

\subsection{Embedding-Space Sequence Compression} \label{subsec:compression}

Single-cell expression profiles are represented over a large fixed gene set, resulting in long input sequences that make full-sequence Transformer modeling computationally expensive. To reduce this cost without changing the token-level modeling objective, we introduced a sequence compression module operating in embedding space. This module shortens the internal sequence processed by the Transformer while preserving the original input and output interfaces of the model.

Let $\mathbf{E}\in\mathbb{R}^{G\times D}$ denote the token embedding sequence for a cell, where $G$ is the number of genes and $D$ is the embedding dimension. At model initialization, we sample and fix a random permutation $\pi$ over gene positions, apply it to $\mathbf{E}$, and partition the reordered sequence into consecutive groups of size $S$, padding at the end if necessary. Writing $G_{\mathrm{c}}=\lceil G/S\rceil$ for the compressed sequence length, each grouped embedding block $\mathbf{E}^{\pi}{(i)}\in\mathbb{R}^{S\times D}$ is flattened and projected to a single $D$-dimensional vector through a shared linear map $W_{\mathrm{down}}\in\mathbb{R}^{D\times (SD)}$:

\begin{equation}
\mathbf{H}_i = W_{\mathrm{down}}\,\mathrm{vec}\!\left(\mathbf{E}^{\pi}_{(i)}\right), \qquad i=1,\dots,G_{\mathrm{c}}, 
\end{equation}
with $\mathbf{E}^\pi = \pi\left(\mathbf{E}\right)$. The resulting compressed sequence $\mathbf{H}=[\mathbf{H}_1,\dots,\mathbf{H}_{G_{\mathrm{c}}}]^\top \in \mathbb{R}^{G_{\mathrm{c}}\times D}$ is then processed by the Transformer backbone of MDDMs.

After the Transformer blocks, we expand the embedding sequence back to full-length before the output projection layer using a shared linear up-projection $W_{\mathrm{up}}\in\mathbb{R}^{(SD)\times D}$. Each compressed representation is mapped back to an $S\times D$ block, concatenated across groups, and reordered with $\pi^{-1}$ to restore the original gene order:

\begin{align}
\widehat{\mathbf{E}}^{\pi}_{(i)} &= \mathrm{unvec}\!\left(W_{\text{up}}\mathbf{H}_i\right), \qquad i=1,\dots,G_{\text{c}}, \\
\widehat{\mathbf{E}} &= \pi^{-1}\!\left(\mathrm{concat}\left(\widehat{\mathbf{E}}^{\pi}_{(1)},\dots,\widehat{\mathbf{E}}^{\pi}_{(G_{\text{c}})}\right)\right).
\end{align}

Thus, compression is applied only to the model's internal computation, while the token-level prediction space remains unchanged. In practice, this substantially reduces computation during training and inference. Moreover, the random grouping with linear projection performs a simple linear mixing of multi-gene signals, thereby reducing the influence of single-gene noise and improving robustness under perturbation.

\subsection{Conditional Generation} \label{subsec:cfg}
\myparagraph{Condition encoding.} For perturbation response prediction, \methodabbr generates single-cell expression token sequences conditioned on a context $c$. In our setting, $c$ consists of two components: a source context specifying the cellular background (for example, a cell line or donor), and a perturbation identity specifying the applied intervention (for example, a target gene or cytokine). Each condition value is represented by a dedicated discrete token appended to the expression-token vocabulary, $\mathcal{V}$. The resulting condition tokens are therefore encoded in the same discrete token space as the expression sequence.

The two condition tokens are prepended to the expression token sequence to form the full model input. These condition tokens are exempt from masking throughout the forward corruption process, so that the conditioning signal remains available at every diffusion step. Under this formulation, the mask predictor is extended from the unconditional setting, $p_\theta\left(x_0|x_t\right)$, to a conditional distribution, $p_\theta\left(x_0|x_t,c\right)$.

\myparagraph{Training process.} Beyond the unconditional model that follows \eqref{eq:loss} to learn $p(\mathbf{x})$, we trained the conditional variant that augments the input with unmasked condition tokens and models $p(\mathbf{x}\mid c)$ using the same objective.

For the conditional model, both perturbed cells and control cells are included during training. Cells without a perturbation-specific target condition are assigned the biologically neutral control label $c_{nt}$ (for example, the \texttt{non-targeting} condition), rather than being treated as missing-condition inputs. This allows the conditional model to learn both perturbation-specific and control-state generation within a single parameterization.

\myparagraph{Sampling with classifier-free guidance (CFG).} At inference time, we use CFG to strengthen perturbation-specific generation. At each denoising step, the conditional model is evaluated under both the target condition $c$ and the control condition $c_{nt}$ as described above.

Let $a_\theta(v \mid x_t, c)$ denote the logits for vocabulary token $v$ at a masked position under condition $c$, and $\tilde{a}_\theta(v \mid x_t, c)$ denote the corresponding guided logits under CFG. The guided logits are computed as
\begin{equation}
\tilde{a}_\theta(v \mid x_t, c) = a_\theta(v \mid x_t, c_{nt}) + (w + 1)\bigl(a_\theta(v \mid x_t, c) - a_\theta(v \mid x_t, c_{nt})\bigr),
\end{equation}
where $w \ge 0$ is the guidance scale. Setting $w=0$ recovers standard conditional generation, whereas larger values of $w$ place greater emphasis on perturbation-specific signal relative to the control condition. The guided logits are converted to sampling probabilities by a softmax at each denoising step.

\subsection{Inference-Time Biological Prior Injection} \label{subsec:prior}

Perturbation response prediction is challenging because perturbation-induced transcriptional changes are often subtle relative to background variability. We therefore introduced a lightweight biological prior at inference time using perturbation profiles from other human cell lines for the target perturbations in the test set. For each perturbation, we constructed a set of downregulated genes $G_{\downarrow}$ consisting of the perturbed target gene together with genes identified as downregulated in individual reference cell lines; the final prior set was obtained by taking the union across available reference cell lines. Details of the data sources and selection criteria are provided in the Appendix.

Sampling was initialized by assigning a low initial expression value $\mu = 1$ to positions corresponding to $g \in G_{\downarrow}$ and leaving all other positions masked:
\begin{equation}
\tilde{x}_g =
\begin{cases}
\mu, & g \in G_{\downarrow},\\
\text{MASK}, & g \notin G_{\downarrow}.
\end{cases}
\end{equation}

We then mapped $\tilde{x}$ to the discrete token space via $q(\cdot)$ to obtain the initial state for reverse diffusion. Positions specified by the biological prior were kept fixed during sampling, while the remaining positions were generated by the model. This strategy provides a directional downregulation signal from other cell lines without assuming direct quantitative transferability across cell systems.

\bibliographystyle{assets/plainnat}
\bibliography{sn-bibliography}

\clearpage
\newpage
\beginappendix

\setcounter{figure}{0}
\setcounter{table}{0}
\setcounter{algorithm}{0}
\renewcommand{\thefigure}{\Alph{section}\arabic{figure}}
\renewcommand{\thetable}{\Alph{section}\arabic{table}}
\renewcommand{\thealgorithm}{\Alph{section}\arabic{algorithm}}

\section{Datasets}\label{sec:datasets}

\subsection{Data processing}

\subsubsection{Data processing for unconditional generation}

\myparagraph{Human datasets.}
Human single-cell RNA-seq data used for unconditional generation were compiled from two sources: the PARSE 10M PBMC dataset~\citep{parse_10m_pbmc_2024} and the CZ CELLxGENE resource~\citep{czi2025cz}. From PARSE, only cells under the PBS control condition were retained. From CZ CELLxGENE, we included cells from eight human tissues: neocortex, lung, breast, thymus, liver, colon, kidney, and heart, restricted to samples annotated as normal, profiled with 10x Genomics 3$'$ v3, and derived from whole cells.

For the human CZ CELLxGENE data, quality control was performed at both the cell and gene levels. Cells were retained if they contained at least 200 detected genes, at least 500 total counts, and a mitochondrial transcript fraction $\leq 20\%$. Genes detected in fewer than 3 cells were excluded. Putative doublets were identified using Scrublet~\citep{wolock2019scrublet} with an expected doublet rate of 0.06, 30 principal components, and a simulated doublet ratio of 2.0, and were removed from downstream analyses.

To harmonize features across all human datasets, gene expression matrices were aligned to a reference set of 18,080 genes used in the Virtual Cell Challenge H1 genetic perturbation dataset~\citep{roohani2025virtual}. For the PARSE dataset, cell level quality control metrics, including total counts, number of detected genes, and mitochondrial transcript fraction, were recalculated after gene alignment. Because PARSE is large in scale and of high overall quality, no additional filtering was applied beyond retaining PBS control cells and aligning genes to the reference set.

\myparagraph{Non-human datasets.}
Non-human single-cell RNA-seq datasets were collected from two sources. Mouse and rhesus macaque datasets were obtained from CZ CELLxGENE~\citep{czi2025cz}, whereas zebrafish and fly datasets were obtained from scBaseCount~\citep{youngblut2025scbasecount}.

For mouse and rhesus macaque, samples were filtered using criteria similar to those applied to the human CZ CELLxGENE data. Specifically, only samples annotated as normal, profiled with 10x Genomics 3$'$ v3, and derived from whole cells were retained. We used mouse ovary data and rhesus macaque lung data in the downstream experiments.

For zebrafish and fly, we retained samples without explicit evidence of disease or perturbation. Specifically, samples were required to have disease annotations indicating the absence of disease, perturbation annotations indicating the absence of treatment or genetic manipulation, library type corresponding to 10x 3$'$ gene expression, and cell preparation annotated as single cell. We used zebrafish embryo data and fly brain data in the downstream experiments.

Quality control for all non-human datasets followed the same general procedure used for the CZ CELLxGENE datasets. Cells were retained if they had at least 200 detected genes, at least 500 total counts, and a mitochondrial transcript fraction no greater than 20\%. Genes detected in fewer than 3 cells were removed. Putative doublets were identified using Scrublet~\citep{wolock2019scrublet} with an expected doublet rate of 0.06, 30 principal components, and a simulated doublet ratio of 2.0, and were excluded prior to downstream analysis.

For non-human species, gene annotations were obtained from Ensembl BioMart~\citep{kinsella2011ensembl} using Ensembl gene identifiers, external gene names, and gene biotypes. Only genes annotated as protein-coding and associated with valid gene names were retained. These annotation tables were used to standardize gene identifiers across datasets and to restrict each dataset to protein-coding genes.

\subsubsection{Data processing for conditional generation}

Conditional generation experiments were conducted in two distinct settings: genetic perturbation generation and cytokine perturbation generation.

\myparagraph{Genetic perturbation datasets.}
For genetic perturbation generation, we used the Virtual Cell Challenge (VCC) H1 cell line dataset~\citep{roohani2025virtual} as the primary benchmark. This dataset contains perturbation targets partitioned into 150 training targets, 50 validation targets, and 100 test targets, enabling evaluation of model generalization to unseen perturbations within the same cellular context.

To increase perturbation diversity during training, we additionally incorporated several publicly available single-cell CRISPR perturbation datasets. These included the genome-wide CRISPRi dataset in K562 and the essential gene perturbation datasets in K562 and RPE1 from Replogle et al.~(2022)~\citep{replogle2022mapping}, the essential-gene Perturb-seq datasets in HepG2 and Jurkat from Nadig et al.~(2025)~\citep{nadig2025transcriptome}, the multi-cell-line Perturb-seq datasets in A549, MCF7, HT29, HAP1, BxPC3, and K562 from Jiang et al.~(2025)~\citep{jiang2025systematic}, and the HCT116 and HEK293T CRISPRi datasets from X-Atlas/Orion~\citep{huang2025x}. From these external datasets, we retained only perturbations whose target genes overlapped with the 300 targets defined in the H1 benchmark, so that the auxiliary training data remained consistent with the target space used for evaluation.

Across all genetic perturbation datasets, gene expression matrices were harmonized to a shared feature space before model training. When combining data from multiple sources, perturbation labels were standardized according to target gene identity, and only cells with unambiguous perturbation annotations were retained for downstream analysis.

\myparagraph{Cytokine perturbation dataset.}
For cytokine perturbation generation, we used the PARSE 10M PBMC cytokine perturbation dataset~\citep{parse_10m_pbmc_2024}, in which perturbation conditions are defined by the cytokine annotation and PBS serves as the control. In this setting, cells with cytokine labels other than PBS were treated as perturbed samples, whereas PBS cells were retained as controls.

Gene expression features were aligned to the same reference set of 18,080 genes used in the unconditional setting. After alignment, cell-level quality control metrics, including total counts, number of detected genes, and mitochondrial transcript fraction, were recalculated. No additional quality control filtering was applied because of the large scale and high overall quality of the dataset.

To evaluate model generalization across both donors and perturbation conditions, the dataset was partitioned into training, validation, and test subsets at the donor level. The training set included all cytokine perturbations from six randomly selected donors, together with all PBS control cells across donors. For validation, two donors were randomly selected; for each donor, 70\% of cytokine perturbation conditions were included in training, whereas the remaining 30\% were held out for validation. PBS cells from these donors were retained as controls. For testing, four additional donors were selected; for each donor, 30\% of cytokine perturbation conditions were included in training, whereas the remaining 70\% were held out for testing. PBS cells from these donors were likewise retained as controls.

\section{Evaluation metrics}\label{sec:metrics}

\subsection{Metrics for unconditional generation}

We evaluate unconditional generation quality using five metrics that assess gene-level expression fidelity and distributional similarity between real and generated cells. Let $\mathbf{X}_{\mathrm{real}} \in \mathbb{R}^{N_r \times G}$ and $\mathbf{X}_{\mathrm{gen}} \in \mathbb{R}^{N_g \times G}$ denote the log1p-normalized expression matrices of $N_r$ real and $N_g$ generated cells over $G$ genes. All metrics are computed on log1p-normalized expression values unless otherwise noted.

\subsubsection{Gene expression correlation}

These metrics quantify how well the generated data recapitulate gene-level expression patterns observed in the real data.

\myparagraph{Pearson correlation.} For each gene $g$, we compute the mean expression across cells in the real and generated populations, yielding mean expression vectors $\bar{\vx}_{\mathrm{real}}, \bar{\vx}_{\mathrm{gen}} \in \mathbb{R}^G$. The Pearson correlation between these two vectors measures gene-level concordance:
\begin{equation}
r_{\mathrm{Pearson}} = \mathrm{corr}\!\left(\bar{\vx}_{\mathrm{real}},\, \bar{\vx}_{\mathrm{gen}}\right).
\end{equation}

\myparagraph{Spearman correlation.} The Spearman rank correlation between the same mean expression vectors, capturing monotonic agreement in gene expression ranks:
\begin{equation}
r_{\mathrm{Spearman}} = \rho_{\mathrm{rank}}\!\left(\bar{\vx}_{\mathrm{real}},\, \bar{\vx}_{\mathrm{gen}}\right).
\end{equation}

\subsubsection{Distributional fidelity}

These metrics assess whether the generated cell population matches the overall distribution of the real data in expression space.

\myparagraph{Maximum Mean Discrepancy (MMD).} MMD measures the distance between two distributions by comparing their mean embeddings in a reproducing kernel Hilbert space. We first compute a joint PCA embedding of real and generated cells and evaluate MMD in this PCA space. Given PCA-projected real samples $\{\vx_i\}_{i=1}^{N_r}$ and generated samples $\{\vy_j\}_{j=1}^{N_g}$ with kernel function $k(\cdot, \cdot)$:
\begin{equation}
\mathrm{MMD}^2 = \frac{1}{N_r^2}\sum_{i,i'}k(\vx_i, \vx_{i'}) - \frac{2}{N_r N_g}\sum_{i,j}k(\vx_i, \vy_j) + \frac{1}{N_g^2}\sum_{j,j'}k(\vy_j, \vy_{j'}).
\end{equation}
We use a multi-scale Gaussian kernel defined as $k(\vx, \vy) = \sum_{s=1}^{S} \exp\!\left(-\|\vx - \vy\|^2 / \sigma_s\right)$, where $S = 5$ scales are constructed by setting a base bandwidth $\sigma_0$ equal to the mean squared pairwise distance, and then generating bandwidths $\sigma_s = \sigma_0 \cdot \alpha^{s - \lceil S/2 \rceil}$ for $s = 1, \dots, S$ with multiplier $\alpha = 2$. Lower values indicate better distributional agreement.

\myparagraph{Gene-averaged 1-Wasserstein distance (1-WD).} We compute the 1-Wasserstein (earth mover's) distance independently for each gene and report the average across all genes:
\begin{equation}
\text{1-WD} = \frac{1}{G}\sum_{g=1}^{G} W_1\!\left(P_{\mathrm{real}}^{(g)},\, P_{\mathrm{gen}}^{(g)}\right),
\end{equation}
where $P_{\mathrm{real}}^{(g)}$ and $P_{\mathrm{gen}}^{(g)}$ are the marginal distributions of gene $g$ in the real and generated populations, respectively. For univariate distributions, $W_1$ reduces to the integral of the absolute difference between the two cumulative distribution functions. Note that this is a per-gene averaged distance, not the multivariate Wasserstein distance over the full $G$-dimensional expression space. Lower values indicate better agreement.

\myparagraph{Integration Local Inverse Simpson's Index (iLISI).} iLISI quantifies how well real and generated cells are mixed in a shared embedding space. For each cell, the local inverse Simpson's index measures the effective number of groups (real vs.\ generated) in its $k$-nearest neighborhood. An iLISI value of 2 indicates perfect mixing, meaning the neighborhood contains equal proportions of real and generated cells; a value of 1 indicates complete separation. We compute iLISI on a joint PCA embedding of real and generated cells, following the procedure of Korsunsky et al.~\citep{korsunsky2019fast}. Higher values indicate better distributional overlap.

\subsection{Metrics for perturbation prediction}
We adopt a panel of seven complementary metrics, implemented in the Cell-Eval toolkit~\citep{adduri2025predicting}, to evaluate perturbation prediction performance. All metrics are computed on pseudobulked expression profiles averaged over cells sharing the same perturbation condition. Let $T$ denote the number of distinct perturbations, $\bar{p}_t$ and $\hat{p}_t$ the observed and predicted pseudobulk expression vectors for perturbation $t$, and $\bar{c}$ the control (unperturbed) pseudobulk. We define expression deltas as $\Delta_t = |\bar{p}_t - \bar{c}|$ (observed) and $\hat{\Delta}_t = |\hat{p}_t - \bar{c}|$ (predicted). Unless stated otherwise, differentially expressed (DE) genes are identified by the Wilcoxon rank-sum test with Benjamini--Hochberg correction at $p_{\mathrm{adj}} < 0.05$.

The seven metrics can be grouped into two categories: \textit{transcriptome-wide accuracy}, which evaluates overall expression-level fidelity across all modeled genes, and \textit{differential expression recovery}, which assesses the identification and ranking of perturbation-responsive genes.

\subsubsection{Transcriptome-wide accuracy}
These metrics assess how closely the predicted pseudobulk profiles match the observed profiles across all genes in terms of expression patterns and perturbation distinguishability.

\myparagraph{Mean Absolute Error (MAE).} The average $\ell_1$ distance between predicted and observed pseudobulks across all perturbations:
\begin{equation}
\mathrm{MAE} = \frac{1}{T}\sum_{t=1}^{T} \|\hat{p}_t - \bar{p}_t\|_1.
\end{equation}

\myparagraph{Pearson Delta Correlation (Pearson-$\Delta$).} The Pearson correlation between predicted and observed expression deltas, computed element-wise across all genes and aggregated over perturbations:
\begin{equation}
\mathrm{Pearson-}\Delta = \mathrm{corr}\!\left(\hat{\Delta}, \Delta\right).
\end{equation}
This metric measures how well the model captures the direction and magnitude of perturbation-induced expression changes at the gene level.

\myparagraph{Perturbation Discrimination Score (PDS).} Adapted from Wu et al.\ (2024)~\citep{wu2024perturbench}, PDS evaluates whether a model can distinguish different perturbation effects. For each perturbation $t$, let $r_t$ be the number of other perturbation ground truths that are closer (in Manhattan distance) to the predicted profile $\hat{p}_t$ than the correct ground truth $\bar{p}_t$. The normalized score is
\begin{equation}
\mathrm{PDS} = 1 - \frac{2}{T}\sum_{t=1}^{T}\frac{r_t}{T},
\end{equation}
where $\mathrm{PDS} = 1$ indicates perfect discrimination and $\mathrm{PDS} = 0$ corresponds to random performance.

\subsubsection{Differential expression recovery}
These metrics evaluate how well the model recovers the set of genes that are significantly differentially expressed in response to each perturbation, as well as the relative effect sizes of different perturbations.

\myparagraph{DE Overlap Accuracy (DES).} For each perturbation, we identify the set of significant DE genes (at $p_{\mathrm{adj}} < 0.05$) in both ground truth and predictions, ranked by absolute log-fold change. Let $\mathcal{G}_{t,\mathrm{true}}^{(N)}$ and $\mathcal{G}_{t,\mathrm{pred}}^{(N)}$ be the top-$N$ DE genes, where $N$ equals the total number of significant DE genes in the ground truth. The overlap accuracy is
\begin{equation}
\mathrm{DES}_t = \frac{|\mathcal{G}_{t,\mathrm{true}}^{(N)} \cap \mathcal{G}_{t,\mathrm{pred}}^{(N)}|}{N}.
\end{equation}

\myparagraph{Spearman Effect Size Correlation (Spearman \#DEG).} The Spearman rank correlation between the number of significant DE genes (DEG) in predictions and ground truth across all perturbations, assessing whether the model captures relative perturbation effect sizes:
\begin{equation}
\mathrm{Spearman\;\#DEG} = \rho_{\mathrm{rank}}\!\left((n_t)_{t=1}^{T},\, (\hat{n}_t)_{t=1}^{T}\right),
\end{equation}
where $n_t = |\mathcal{G}_{t,\mathrm{true}}^{(\mathrm{DE})}|$ and $\hat{n}_t = |\mathcal{G}_{t,\mathrm{pred}}^{(\mathrm{DE})}|$.

\myparagraph{Spearman Log-Fold-Change Correlation (Spearman LFC).} The Spearman rank correlation between predicted and observed log-fold changes, restricted to genes that are significantly DE in the ground truth:
\begin{equation}
\mathrm{Spearman\;LFC}_t = \rho_{\mathrm{rank}}\!\left(\hat{\Delta}_{t,\mathcal{G}_t^*},\, \Delta_{t,\mathcal{G}_t^*}\right),
\end{equation}
where $\mathcal{G}_t^*$ is the set of significant DE genes for perturbation $t$.

\myparagraph{Precision--Recall AUC (AUPRC).} To measure the model's ability to identify significant DE genes with high precision, we assign binary labels to each gene (1 if $p_{\mathrm{adj}} < 0.05$ in ground truth, 0 otherwise) and use predicted $-\!\log_{10}(p_{\mathrm{adj}})$ values as confidence scores. AUPRC is the area under the precision--recall curve:
\begin{equation}
\mathrm{AUPRC}_t = \int_0^1 \mathrm{Precision}_t(r)\,\mathrm{d}\;\mathrm{Recall}.
\end{equation}

\myparagraph{Average score.} To summarize overall performance across the seven perturbation-prediction metrics, we define an average score that measures how much a method improves over the cell-mean baseline, relative to the baseline's remaining room for improvement. Here, the cell-mean baseline predicts each perturbation as the control pseudobulk, i.e., $\hat{p}_t=\bar{c}$ for all $t$.

For each metric $m \in \{\text{MAE}, \text{Pearson-}\Delta, \text{PDS}, \text{DES}, \text{Spearman \#DEG}, \text{Spearman LFC}, \text{AUPRC}\}$, let $s^{(m)}_{\text{method}}$ and $s^{(m)}_{\text{base}}$ denote the score of the evaluated method and the cell-mean baseline, respectively. We define the relative improvement for metric $m$ as
\begin{equation}
r^{(m)} =
\begin{cases}
\max\left\{\dfrac{s^{(m)}_{\text{base}} - s^{(m)}_{\text{method}}}{s^{(m)}_{\text{base}}},\, 0\right\}, & m=\text{MAE},\\[1.2ex]
\max\left\{\dfrac{s^{(m)}_{\text{method}} - s^{(m)}_{\text{base}}}{1 - s^{(m)}_{\text{base}}},\, 0\right\}, & \text{otherwise},
\end{cases}
\end{equation}
where MAE is smaller-is-better, while the other six metrics are larger-is-better and upper-bounded by $1$. The average score is the arithmetic mean of these seven relative improvements:
\begin{equation}
\mathrm{Average\ score} = \frac{1}{7}\sum_{m} r^{(m)}.
\end{equation}
A score of zero indicates no improvement over the cell-mean baseline on any metric, while a score of one indicates perfect performance on all metrics.

\subsection{Full VCC generalist leaderboard}\label{subsec:vcc-leaderboard}

Table~\ref{tab:vcc-leaderboard-full} reports detailed results for all 26 entries on the Virtual Cell Challenge (VCC) generalist leaderboard, including \methodabbr{} evaluated post-hoc. For each of the seven perturbation-prediction metrics, we list both the raw value and the per-metric rank across all entries. Entries are sorted by average rank (ascending). \methodabbr{} achieves the best average rank (8.7) and ranks first on two of seven metrics (MAE and Pearson-$\Delta$).

\begin{table}[t]
\centering
\caption{Full VCC generalist track leaderboard. Bold values indicate the best performance for each metric. \methodabbr{} was evaluated post-hoc and is separated by a horizontal rule.
For each metric, ``val'' denotes the metric value and ``rk'' denotes the per-metric rank across all 26 entries.
Abbreviated metric names: Sp. \#DEG = Spearman \#DEG; Sp.~LFC = Spearman LFC; Pear.-$\Delta$ = Pearson-$\Delta$. Metric definitions are given in Section~\ref{sec:metrics}. Original leaderboard data are available at \url{https://virtualcellchallenge.org/leaderboard}.}
\label{tab:vcc-leaderboard-full}
\footnotesize
\setlength{\tabcolsep}{2pt}
\begin{tabular*}{\textwidth}{@{\extracolsep\fill}lr|rr|rr|rr|rr|rr|rr|rr}
\toprule
 & & \multicolumn{2}{c}{DES$\uparrow$} & \multicolumn{2}{c}{PDS$\uparrow$} & \multicolumn{2}{c}{MAE$\downarrow$} & \multicolumn{2}{c}{Sp. \#DEG$\uparrow$} & \multicolumn{2}{c}{Sp.\ LFC$\uparrow$} & \multicolumn{2}{c}{AUPRC$\uparrow$} & \multicolumn{2}{c}{Pear.-$\Delta$$\uparrow$} \\
\cmidrule(lr){3-4}\cmidrule(lr){5-6}\cmidrule(lr){7-8}\cmidrule(lr){9-10}\cmidrule(lr){11-12}\cmidrule(lr){13-14}\cmidrule(lr){15-16}
Team & Avg Rank$\downarrow$ & val & rk & val & rk & val & rk & val & rk & val & rk & val & rk & val & rk \\
\midrule
\methodabbr{} & \textbf{8.7} & 0.216 & 25 & 0.748 & 9 & \textbf{0.052} & 1 & 0.394 & 12 & 0.331 & 3 & 0.272 & 10 & \textbf{0.306} & 1 \\
\midrule
cleopatra & 9.1 & 0.228 & 24 & 0.747 & 10 & 0.086 & 3 & 0.473 & 9 & \textbf{0.396} & 1 & 0.266 & 11 & 0.203 & 6 \\
xBio & 10.7 & 0.305 & 13 & 0.811 & 6 & 0.770 & 13 & 0.564 & 4 & 0.087 & 20 & 0.252 & 16 & 0.217 & 3 \\
Cellock Holmes & 10.9 & 0.356 & 3 & 0.679 & 26 & 0.239 & 7 & 0.000 & 19 & 0.238 & 6 & 0.576 & 4 & 0.125 & 11 \\
Shippers & 11.0 & 0.354 & 6 & 0.699 & 22 & 0.231 & 6 & 0.000 & 19 & 0.227 & 8 & 0.576 & 4 & 0.123 & 12 \\
Mean Predictors & 11.4 & 0.305 & 13 & 0.741 & 11 & 6.72 & 26 & 0.294 & 15 & 0.213 & 11 & \textbf{0.582} & 1 & 0.217 & 3 \\
rdbs & 12.1 & 0.332 & 8 & 0.696 & 24 & 0.536 & 10 & 0.000 & 19 & 0.233 & 7 & 0.576 & 4 & 0.119 & 13 \\
WIND & 12.7 & 0.305 & 13 & 0.735 & 13 & 0.621 & 11 & 0.141 & 16 & 0.280 & 4 & 0.165 & 22 & 0.127 & 10 \\
vegansnail & 12.9 & 0.330 & 9 & 0.687 & 25 & 0.526 & 9 & 0.000 & 19 & 0.222 & 10 & 0.576 & 4 & 0.117 & 14 \\
Nathan LaPierre & 13.1 & 0.246 & 21 & 0.710 & 18 & 0.121 & 4 & 0.444 & 10 & 0.114 & 18 & 0.225 & 19 & 0.218 & 2 \\
Outlier & 13.3 & 0.361 & 2 & 0.845 & 3 & 4.22 & 25 & 0.354 & 13 & 0.083 & 21 & 0.572 & 8 & 0.098 & 21 \\
BM\_xTVCBM & 13.3 & 0.349 & 7 & \textbf{0.872} & 1 & 1.03 & 15 & 0.000 & 19 & 0.046 & 24 & \textbf{0.582} & 1 & 0.045 & 26 \\
LIUtest & 13.6 & 0.273 & 19 & 0.698 & 23 & 0.084 & 2 & 0.305 & 14 & 0.208 & 12 & 0.245 & 18 & 0.187 & 7 \\
BioCai & 13.7 & \textbf{0.366} & 1 & 0.799 & 7 & 1.45 & 18 & 0.000 & 19 & 0.045 & 25 & \textbf{0.582} & 1 & 0.047 & 25 \\
XLearning Lab & 13.9 & 0.356 & 3 & 0.851 & 2 & 1.07 & 16 & 0.042 & 18 & 0.056 & 23 & 0.263 & 12 & 0.055 & 23 \\
NetPhar & 14.1 & 0.269 & 20 & 0.736 & 12 & 1.53 & 19 & 0.582 & 3 & 0.129 & 16 & 0.262 & 13 & 0.109 & 16 \\
Xcompass & 14.3 & 0.318 & 10 & 0.812 & 5 & 4.21 & 24 & 0.135 & 17 & 0.108 & 19 & 0.182 & 20 & 0.212 & 5 \\
Turtle & 14.6 & 0.239 & 23 & 0.723 & 15 & 2.18 & 22 & 0.669 & 2 & 0.239 & 5 & 0.261 & 15 & 0.099 & 20 \\
xinheng & 14.7 & 0.310 & 11 & 0.702 & 21 & 0.687 & 12 & 0.511 & 6 & 0.139 & 15 & 0.165 & 22 & 0.109 & 16 \\
4vcc & 14.7 & 0.121 & 26 & 0.760 & 8 & 0.124 & 5 & -0.112 & 26 & 0.367 & 2 & 0.166 & 21 & 0.113 & 15 \\
Gavin & 15.0 & 0.241 & 22 & 0.721 & 16 & 2.18 & 22 & \textbf{0.684} & 1 & 0.226 & 9 & 0.262 & 13 & 0.093 & 22 \\
SIMON & 15.0 & 0.355 & 5 & 0.818 & 4 & 1.11 & 17 & -0.049 & 25 & 0.059 & 22 & 0.278 & 9 & 0.055 & 23 \\
VCC & 15.1 & 0.303 & 16 & 0.703 & 20 & 0.409 & 8 & 0.511 & 6 & 0.141 & 14 & 0.164 & 24 & 0.107 & 18 \\
TF\_Boys & 15.9 & 0.296 & 17 & 0.719 & 17 & 1.73 & 21 & 0.443 & 11 & 0.163 & 13 & 0.164 & 24 & 0.161 & 8 \\
Anyone & 16.0 & 0.309 & 12 & 0.705 & 19 & 0.835 & 14 & 0.562 & 5 & 0.124 & 17 & 0.163 & 26 & 0.106 & 19 \\
dpdg3157 & 16.0 & 0.294 & 18 & 0.730 & 14 & 1.67 & 20 & 0.506 & 8 & 0.041 & 26 & 0.249 & 17 & 0.156 & 9 \\
\botrule
\end{tabular*}

\end{table}

\section{Details on \methodabbr{}}\label{sec:method-details}

\subsection{Model architecture}\label{subsec:architecture}

\begin{figure}[t]
\centering
\includegraphics[width=\linewidth]{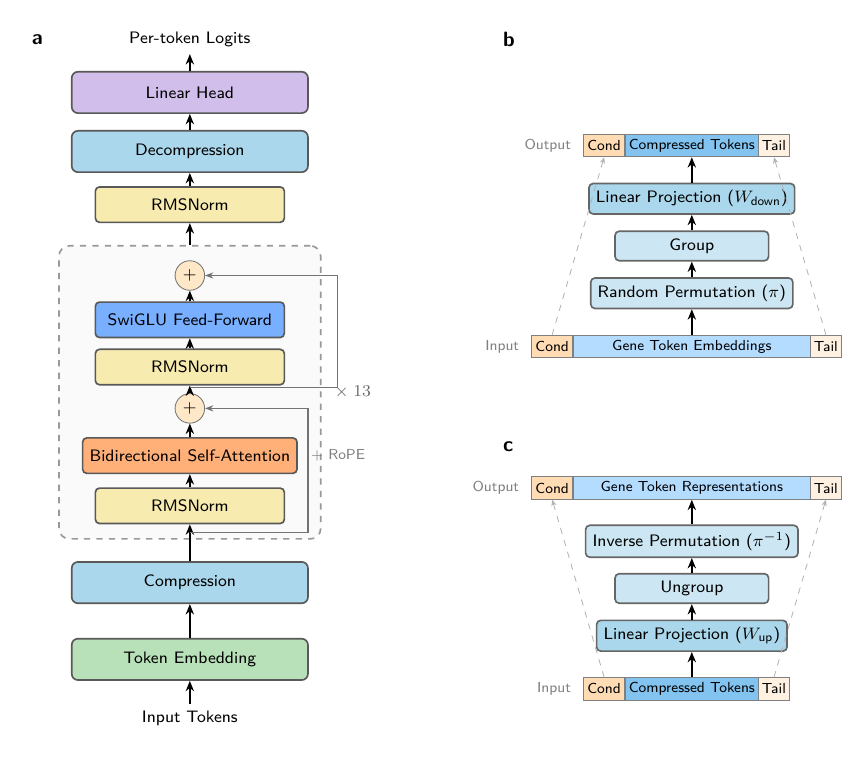}
\caption{\textbf{Model architecture of \methodabbr{}.}
\textbf{a},~Overall model network architecture. Input tokens are first embedded and then passed through a compression module that reduces the gene token sequence length. The compressed sequence is processed by a stack of 13 Transformer blocks, each consisting of RMSNorm, bidirectional self-attention with rotary position embeddings (RoPE), and a SwiGLU feed-forward network, with residual connections around each sub-layer. A final RMSNorm is applied before the decompression module restores the original gene-level resolution, and a linear head produces per-token logits over the expression vocabulary.
\textbf{b},~Compression module. Gene token embeddings are reordered with a fixed random permutation~$\pi$, partitioned into consecutive groups, and linearly projected via~$W_{\text{down}}$ to produce a shorter sequence of compressed tokens. Condition prefix and tail tokens bypass this module.
\textbf{c},~Decompression module. The inverse of compression: compressed tokens are linearly projected via~$W_{\text{up}}$, ungrouped, and reordered with the inverse permutation~$\pi^{-1}$ to recover gene-level representations.}
\label{fig:architecture}
\end{figure}

\begin{figure}[t]
\centering
\includegraphics[width=\linewidth]{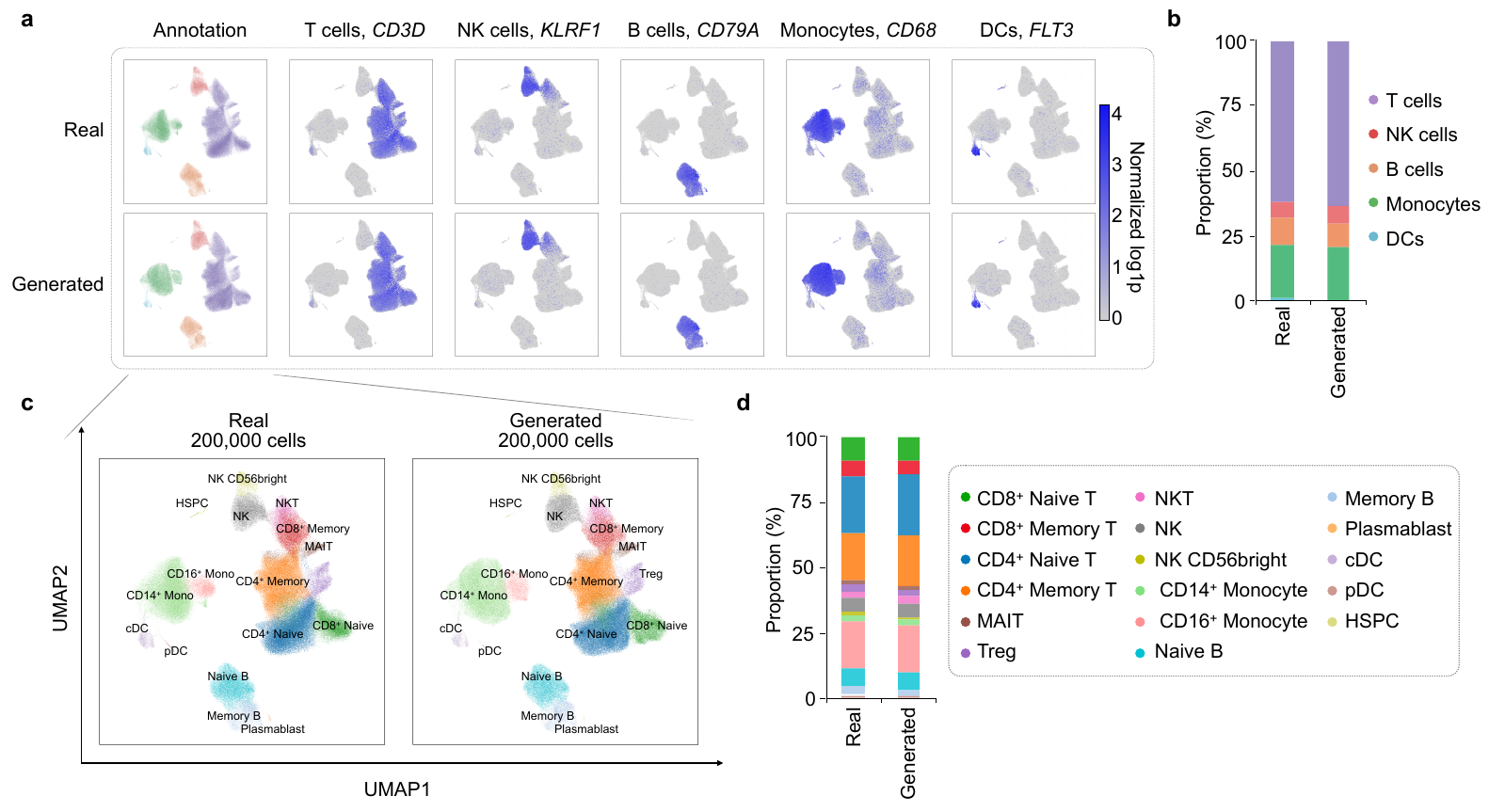}
\caption{\textbf{Scaling unconditional generation to larger cell populations.}
\textbf{a}, UMAP visualization of real and generated cells from the PARSE 10M PBMC dataset, colored by cell type annotation (left) and normalized expression (log1p) of canonical marker genes.
\textbf{b}, Comparison of major cell type proportions between real and generated data.
\textbf{c}, UMAP visualization at higher resolution, showing that generated cells accurately recapitulate cell subtype structure of real data.
\textbf{d}, Comparison of cell subtype proportions between real and generated data, demonstrating robust consistency at higher resolution.}
\label{fig:ext-data-fig1}
\end{figure}

\begin{figure}[t]
\centering
\includegraphics[width=0.8\linewidth]{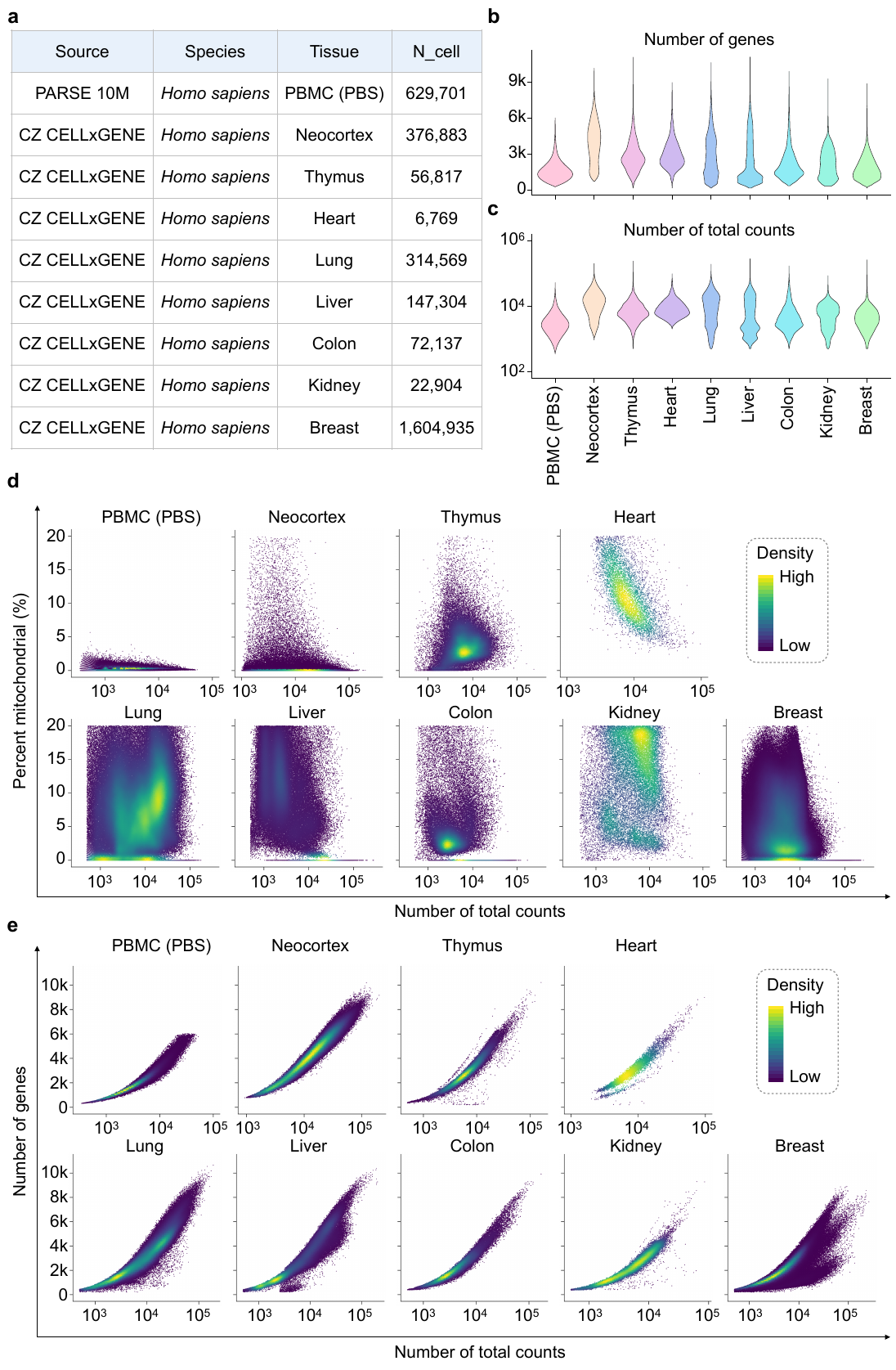}
\caption{\textbf{Quality control of human unconditional generation datasets.}
\textbf{a}, Summary of datasets used for unconditional generation, including data source, species, tissue type and cell number.
\textbf{b},\textbf{c}, Distribution of the number of detected genes (\textbf{b}) and total counts (\textbf{c}) across tissues.
\textbf{d}, Density scatter plots of mitochondrial gene percentage versus total counts for each tissue.
\textbf{e}, Density scatter plots of the number of detected genes versus total counts for each tissue.}
\label{fig:ext-data-fig2}
\end{figure}

\begin{figure}[t]
\centering
\includegraphics[width=\linewidth]{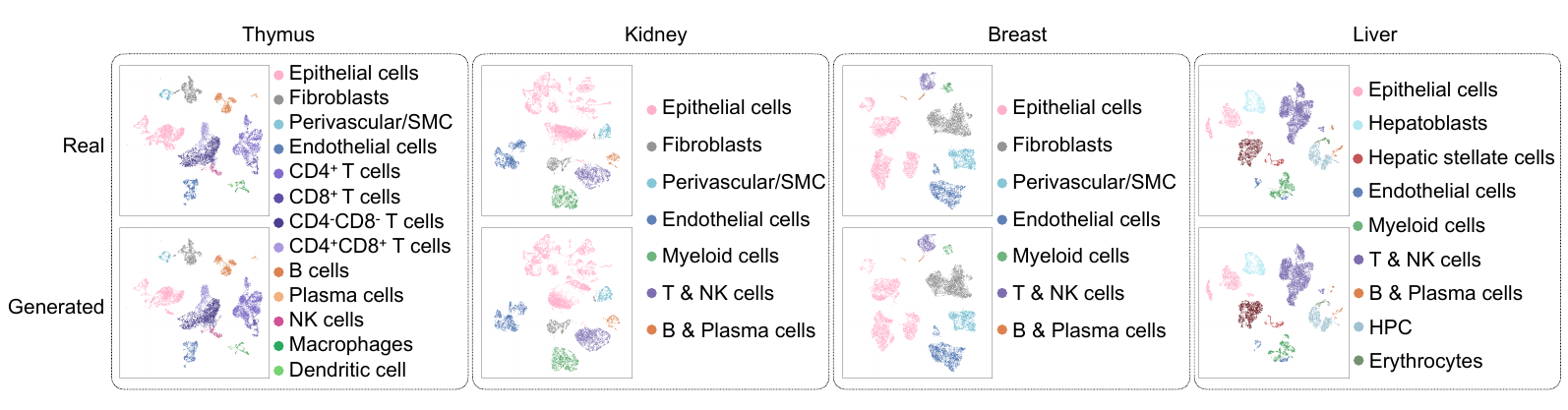}
\caption{\textbf{Unconditional generation results for additional human tissues.}
UMAP visualization of real (top) and generated (bottom) cells for thymus, kidney, breast and liver, colored by cell type annotation.}
\label{fig:ext-data-fig3}
\end{figure}

\begin{figure}[t]
\centering
\includegraphics[width=\linewidth]{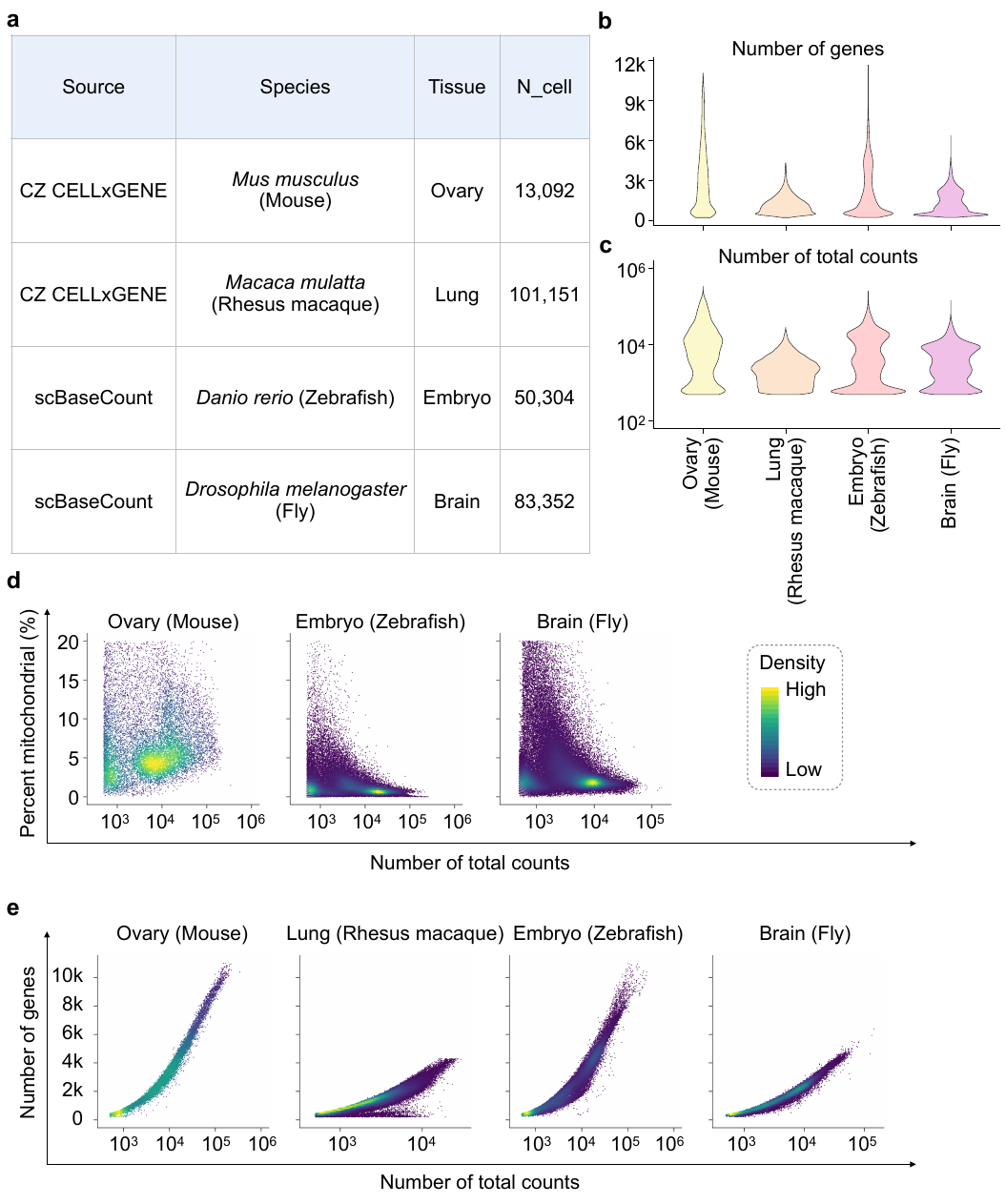}
\caption{\textbf{Quality control of non-human species unconditional generation datasets.}
\textbf{a}, Summary of datasets used for unconditional generation across non-human species, including data source, species, tissue type and cell number.
\textbf{b},\textbf{c}, Distribution of the number of detected genes (\textbf{b}) and total counts (\textbf{c}) across species.
\textbf{d}, Density scatter plots of mitochondrial gene percentage versus total counts for each species. The rhesus macaque lung dataset is not shown owing to the absence of annotated mitochondrial genes.
\textbf{e}, Density scatter plots of the number of detected genes versus total counts for each species.}
\label{fig:ext-data-fig4}
\end{figure}

The mask predictor network of \methodabbr{} is a bidirectional Transformer operating on compressed token sequences (Figure~\ref{fig:architecture}).
Given an input sequence of $G=18{,}080$ gene expression tokens plus a short condition prefix, the model first maps all tokens to $D$-dimensional embeddings through a shared embedding layer.
The gene token embeddings are then compressed from length $G$ to $G_{\mathrm{c}} = \lceil G/S \rceil$ using the sequence compression module described in Section~\ref{subsec:compression}. The default group size is $S = 8$ ($G_{\mathrm{c}} = 2{,}260$); for genetic perturbation prediction, $S = 32$ ($G_{\mathrm{c}} = 565$) is used, as ablation experiments showed improved prediction performance at this compression ratio in addition to reduced computational cost.
The condition prefix tokens bypass compression and are concatenated with the compressed tokens before entering the Transformer backbone.

The Transformer backbone follows a LLaMA-style architecture and consists of $L = 13$ identical blocks.
Each block applies two sub-layers with pre-normalization and residual connections:
\begin{equation}
\vx \leftarrow \vx + \mathrm{Attn}\!\left(\mathrm{RMSNorm}(\vx)\right), \qquad
\vx \leftarrow \vx + \mathrm{FFN}\!\left(\mathrm{RMSNorm}(\vx)\right).
\end{equation}
The self-attention sub-layer uses multi-head attention with $n_{\mathrm{h}} = 10$ heads (head dimension $d_{\mathrm{h}} = 64$) and applies rotary position embeddings (RoPE) to the query and key vectors in every block.
Attention is bidirectional (no causal mask), consistent with the order-independent nature of masked diffusion models.
The feed-forward sub-layer uses a SwiGLU activation with an intermediate dimension of $d_{\mathrm{ff}} = 2{,}560$.
All linear layers are bias-free.

After the final Transformer block, a final RMSNorm is applied, and the decompression module (inverse of compression; Section~\ref{subsec:compression}) restores the sequence to the original gene-level length.
A linear output head then maps each position to logits over the expression vocabulary.

Table~\ref{tab:architecture} summarizes the key architectural hyperparameters.

\begin{table}[t]
\centering
\caption{Architectural hyperparameters of \methodabbr{}.}
\label{tab:architecture}
\begin{tabular}{ll}
\toprule
Hyperparameter & Value \\
\midrule
Embedding dimension ($D$) & 640 \\
Transformer blocks ($L$) & 13 \\
Attention heads ($n_{\mathrm{h}}$) & 10 \\
Head dimension ($d_{\mathrm{h}}$) & 64 \\
Feed-forward dimension ($d_{\mathrm{ff}}$) & 2{,}560 \\
Position encoding & RoPE \\
Attention type & Bidirectional \\
Normalization & RMSNorm \\
Activation & SwiGLU \\
Compression group size ($S$) & 8 (32 for genetic perturbation prediction) \\
Gene sequence length ($G$) & 18{,}080 \\
\bottomrule
\end{tabular}
\end{table}

\begin{table}[t]
\centering
\caption{Training configurations for different experimental settings.}
\label{tab:training-details}
\begin{tabular}{lcccc}
\toprule
& \multicolumn{2}{c}{Unconditional} & \multicolumn{2}{c}{Conditional} \\
\cmidrule(lr){2-3} \cmidrule(lr){4-5}
Setting & PARSE & Per-tissue & \makecell{Genetic \\ perturbation} & \makecell{Cytokine \\ perturbation} \\
\midrule
Dataset & PARSE PBS & \makecell{CZ CELLxGENE \\ scBaseCount} & \makecell{VCC H1 \\ + auxiliary} & PARSE 10M \\
GPUs & $8$ & $2$--$4$ & $16$ & $8$ \\
Peak LR & $1 \times 10^{-4}$ & $2 \times 10^{-4}$ & $2 \times 10^{-4}$ & $2 \times 10^{-4}$ \\
Warmup steps & 100 & 100 & 1{,}000 & 1{,}000 \\
\bottomrule
\end{tabular}

\end{table}

\begin{table}[!htbp]
\centering
\caption{Training data composition for conditional generation tasks.}\label{tab:training-data-stats}
\begin{tabular}{llrrr}
\toprule
Task & Dataset & \#Perturbed cells & \makecell{\#Original \\ control cells} &  \makecell{\#Sub-sampled \\ control cells} \\
\midrule
\multirow{2}{*}{Genetic perturbation} & H1 & 221,273 & 38,176 & 20,344 \\
& External & 323,913 & 583,377 & 35,990 \\
\midrule
Cytokine perturbation & PARSE 10M & 6,499,077 & 629,701 & 629,701 \\
\bottomrule
\end{tabular}

\end{table}

\subsection{Training and inference details}\label{subsec:training-details}

\myparagraph{Training.}
All models were trained using AdamW~\citep{loshchilov2017decoupled} with $\beta_1 = 0.9$, $\beta_2 = 0.95$, and weight decay $0.1$. Gradients were clipped to a maximum norm of $1.0$. The learning rate followed a cosine annealing schedule with linear warmup, decaying from the peak value to a minimum of $2 \times 10^{-5}$. Training was performed in bfloat16 mixed precision with distributed data parallel (DDP) on NVIDIA A800 GPUs. We maintained two exponential moving average (EMA) copies of the model weights using the power-function EMA schedule~\citep{karras2024analyzing} with relative standard deviations $\sigma_{\mathrm{rel}} \in \{0.050, 0.100\}$; all reported results use the $\sigma_{\mathrm{rel}} = 0.050$ checkpoint.

The global batch size was 256 for all settings. \Cref{tab:training-details} summarizes additional training configurations. For genetic perturbation prediction, the model was trained on the combined H1 training set and external perturbation datasets (\Cref{sec:datasets}), with the proportion of control cells capped at $10\%$. The same control ratio was used for cytokine perturbation prediction on the PARSE 10M dataset. The resulting training set sizes are summarized in Table~\ref{tab:training-data-stats}. Unconditional models were trained independently per tissue or species dataset.

\myparagraph{Inference.}
We used a cosine timestep schedule to discretize the reverse diffusion process into $N$ steps. Before sampling begins, the number of tokens to unmask at each step was precomputed and fixed from this schedule. Within each step, the positions to unmask were selected uniformly at random from the remaining masked tokens, and the token values at those positions were drawn from the model's predicted categorical distribution without temperature scaling or top-$k$ truncation.
For unconditional generation, we used $N = 256$ steps without CFG.
For conditional generation, we used $N = 3$ steps with classifier-free guidance (CFG); the guidance weight was set to $w = 2$ for genetic perturbation prediction and $w = 3$ for cytokine perturbation prediction.

\subsection{Identification of perturbation-specific prior genes}

To introduce biologically informed priors into conditional generation, we derived perturbation-specific prior genes from external single-cell CRISPR perturbation datasets using the \texttt{pdex} Python package~\citep{adduri2025predicting}. For each external dataset, differential expression analysis was performed by comparing each perturbation group against the corresponding non-targeting control group across the shared 18,080-gene feature space. We used the default differential expression procedure implemented in \texttt{pdex}, which computes per-gene statistics using Mann--Whitney U tests followed by false discovery rate (FDR) correction.

For each perturbation target and each gene, \texttt{pdex} reports the mean expression in perturbed cells (\texttt{target\_mean}) and the mean expression in the corresponding non-targeting control cells (\texttt{reference\_mean}), together with fold change and FDR. We defined perturbation-specific prior genes as those satisfying
\begin{align}
\mathrm{FDR} &< 0.05, \\
|\mathrm{target\_mean} - \mathrm{reference\_mean}| &> 1, \\
|\log_2(\mathrm{fold\_change})| &> 1, \\
\mathrm{reference\_mean} &> 2, \\
\mathrm{target\_mean} &< 1.
\end{align}

Thus, retained genes were significantly differentially expressed, showed large effect sizes, and were strongly downregulated from the non-targeting control state to the perturbed state.

When the same perturbation target was observed in multiple external cell lines, the corresponding prior gene sets were merged by taking their union across cell lines to construct a unified perturbation-specific prior. During conditional generation, these prior genes were preferentially used to initialize masked positions at the start of generation, thereby injecting biologically informed perturbation signals into the sampling process.

\end{document}